\documentclass[a4paper,11pt]{article}
\usepackage[utf8]{inputenc}
\usepackage[margin=1.0in]{geometry}
\usepackage{amsmath,amssymb,amsthm}
\usepackage[shortlabels]{enumitem}
\usepackage[hidelinks]{hyperref}
\usepackage[noabbrev,capitalize,nameinlink]{cleveref}

\usepackage{natbib}

\usepackage{newpxtext,newpxmath}
\usepackage{bbm}
\usepackage{setspace}

\theoremstyle{plain}
\newtheorem{theorem}{Theorem}
\newtheorem{proposition}{Proposition}
\newtheorem{lemma}{Lemma}
\newtheorem{corollary}{Corollary}

\newtheorem*{mastertheorem}{Master Theorem}

\newcounter{mastertheorem}

\crefname{mastertheorem}{Master Theorem}{Master Theorems}
\Crefname{mastertheorem}{Master Theorem}{Master Theorems}

\crefformat{mastertheorem}{#2Master Theorem#3}
\Crefformat{mastertheorem}{#2Master Theorem#3}

\theoremstyle{definition}
\newtheorem*{definition*}{Definition}

\theoremstyle{remark}
\newtheorem{remark}{Remark}
\newtheorem*{remark*}{Remark}

\usepackage{apptools}
\AtAppendix{\counterwithin{theorem}{section}}
\AtAppendix{\counterwithin{lemma}{section}}
\AtAppendix{\counterwithin{corollary}{section}}
\AtAppendix{\counterwithin{proposition}{section}}

\usepackage{manyfoot}
\SetFootnoteHook{\hspace*{-1.8em}}
\DeclareNewFootnote{bl}[gobble]
\newcommand{\Poisson}{\operatorname{Poisson}}
\newcommand{\actionvec}{\mathbf{a}}
\newcommand{\m}{\mathbf m}

\begin{document}

\title{\textsc{On the existence of pure epsilon-equilibria}}
\author{Bary S.R. Pradelski\thanks{CNRS, Maison Fran\c{c}aise d'Oxford; Department of Economics, University of Oxford} \and Bassel Tarbush\thanks{Merton College and Department of Economics, University of Oxford}}
\date{\today}

\maketitle

\begin{abstract} 
\noindent
Departing from optimizing behavior by any $\epsilon>0$ ensures the existence of stable outcomes in a broad class of many-agent games where utilities are drawn from a joint distribution. Specifically, in a game with iid drawn utilities from a continuous distribution and any $\epsilon>0$, we show that as the number of agents gets large the probability that a pure $\epsilon$-equilibrium exists converges to 1, while, by contrast, the probability that a pure Nash equilibrium exists is known to converge to  $1-1/e\approx 0.63$. The asymptotic existence of pure $\epsilon$-equilibrium extends to utilities that are negatively correlated across agents, and we identify a correlation threshold dependent on $\epsilon$ that is tight for the zero-sum case. We also study the cases of correlation across action profiles and of utility ties and, in all cases, we also characterize the asymptotic distribution of pure $\epsilon$-equilibria. Regarding pure Nash equilibrium---i.e., when $\epsilon=0$---our results nest, organize, and extend the existing literature. All our results and those from the prior literature are applications of our ``master theorem'', which builds on the Chen--Stein method and Stein's method.
\end{abstract}
\footnotebl{\emph{Email}: \texttt{bary.pradelski@cnrs.fr}, \texttt{bassel.tarbush@economics.ox.ac.uk}}
\footnotebl{\emph{Keywords}: pure epsilon-equilibrium, pure Nash equilibrium, equilibrium existence}
\footnotebl{\emph{Thanks}:  We are grateful for comments and suggestions from Simon Jantschgi, Edwin Lock, Patrick Loiseau, Panayotis Mertikopoulos, Jonny Newton, John Quah, Marco Scarsini, Eran Shmaya, Ludvig Sinander, Alex Teytelboym, and Peyton Young.}

\section{Introduction}

In a pure $\epsilon$-equilibrium agents play actions that yield a utility that is within $\epsilon$ of the utility that results from best responding given the actions of others. Unlike Nash equilibrium, $\epsilon$-equilibrium relaxes the requirement that agents fully optimize, which is consistent with the literature on bounded rationality and is especially pertinent in complex or large interactive settings.
While a pure Nash equilibrium may not exist in every game,\footnote{Existence is guaranteed only if mixed strategies are allowed \citep{nash1950equilibrium}, which may lack foundations in some settings \citep[cf. discussion in][]{rubinstein1991comments}. Pure Nash equilibrium is guaranteed to exist in certain classes of games; e.g., in potential games \citep{monderer1996potential} and in supermodular games \citep{milgrom1990rationalizability}.}
 a pure $\epsilon$-equilibrium exists trivially in any game for sufficiently large $\epsilon$. 
This raises a natural question:
\begin{center}
 For which values of $\epsilon \geq 0$ does pure $\epsilon$-equilibrium exist?
 \end{center}

In general, the answer to this question depends on the specific game at hand; for example, the matching pennies game has no pure $\epsilon$-equilibrium for $\epsilon<2$, but every outcome is a pure $\epsilon$-equilibrium for $\epsilon \geq 2$. 

In this paper, we go beyond specific games by endowing the space of all games with $n$ agents and a uniformly bounded number of actions with a probability measure that results from utilities being drawn according to some joint distribution, and we study its properties under the large-$n$ limit.\footnote{We focus on the limit in the number of agents, but the limit in the number(s) of actions may also be of interest, and has been studied in the literature for pure Nash equilibria. A previous version of this article also contained results about the action limit (cf. \url{https://arxiv.org/abs/2502.07585v3}).} 
This approach has a long tradition in the study of the existence and distribution of pure Nash equilibria under a variety of probability measures. In particular, utilities being drawn 
(i) from a continuous distribution (which results in no ties) and independently across agents and action profiles \citep*[cf., e.g.,][]{goldberg1968probability,arratia1990poisson}, 
(ii) 
independently across action profiles but with positive or negative correlation across agents \citep*{amiet2021pure},
(iii)  independently across agents but positively correlated across action profiles \citep*{daskalakis2011connectivity}, or
(iv) independently across agents and action profiles, but from a discrete distribution (which results in ties) \citep*{powers1990limiting,amiet2021pure}. 

Our first set of contributions shows that---while pure Nash equilibria may fail to exist---departing from optimizing behavior by any $\epsilon >0$  ensures the existence of pure $\epsilon$-equilibria in a broad class of many-agent games. Our second set of contributions regards pure Nash equilibrium---i.e., the special case when setting $\epsilon=0$---for which our results nest, organize, and extend the existing literature. In particular: 

\begin{enumerate}[(i),topsep=0ex,itemsep=0ex,leftmargin=0cm]
\item When utilities are drawn independently across agents and action profiles from a continuous distribution (which results in no ties), then, as the number of agents gets large, the probability that a pure Nash equilibrium exists converges to $1-1/e\approx 0.63$ and their number converges in distribution to Poisson(1) (cf. \citealt {arratia1990poisson}, which we recover in \cref{thm:independent_continuous} Part (a)). By contrast, in \cref{thm:independent_continuous} Part (b), we show that for \emph{any} $\epsilon>0$, the probability that a pure $\epsilon$-equilibrium exists converges to 1 and, provided $\epsilon$ is smaller than the range of possible utilities, their standardized number converges in distribution to a standard Gaussian.
\item To examine the robustness of our benchmark results under independence, we study the case where  utilities are drawn independently across action profiles but with positive or negative correlation across agents. Then, as the number of agents gets large, the probability that a pure Nash equilibrium exists converges to 1 and their standardized number converges in distribution to a standard Gaussian (\cref{thm:correlation} Part (a)).
Moreover,  in \cref{thm:correlation} Part (b), we show that, for \emph{any} $\epsilon>0$, the probability that a pure $\epsilon$-equilibrium exists converges to 1 provided the correlation-tuning parameter is bounded below by a constant exceeding $-2\epsilon$.  Thus, for $\epsilon>0$, existence obtains even for negatively correlated utilities. Moreover, their standardized number converges in distribution to a standard Gaussian provided $\epsilon$ is smaller than the range of possible utilities. For zero-sum games---i.e., maximal negative correlation---in which agents have sufficiently many actions, the probability that a pure $\epsilon$-equilibrium with $\epsilon<1/2$ exists converges to 0 (\cref{thm:correlation-nonexistence}). These results extend those of \cite{Rin00} for pure Nash equilibrium.
\item Next, we consider the other dimension of dependence, where utilities are drawn independently across agents but are correlated across action profiles.  For this case, we follow and extend  \cite{daskalakis2011connectivity}, and consider positive correlation across action profiles, modeled via graphical games on interaction graphs with sufficiently strong (logarithmic) expansion.  Then, as the number of agents gets large, the number of pure Nash equilibria converges in distribution to Poisson(1) and hence the probability that a pure Nash equilibrium exists converges to $1-1/e$ (\cref{thm:graphical} Part (a)). 
Next, for $\epsilon>0$, the standardized number of pure $\epsilon$-equilibria converges in distribution to a standard Gaussian and the probability that a pure $\epsilon$-equilibrium exists converges to $1$ (\cref{thm:graphical} Part (b)).
\item Finally, we consider utility ties under the full independence model, which arise naturally when utilities are measured at a finite resolution---intuitively, playing a similar role to the tolerance of $\epsilon>0$. As the number of agents gets large, if the set from which utilities are drawn grows faster than the number of agents then the number of pure Nash equilibria converges in distribution to Poisson(1) and hence the probability that a pure Nash equilibrium exists converges to $1-1/e$ (\cref{thm:independent_discrete} Part (a)), and, if the set from which utilities are drawn grows slower than the number of agents, then the standardized number of pure Nash equilibria converges in distribution to a standard Gaussian and the probability that a pure Nash equilibrium exists converges to 1 (\cref{thm:independent_discrete}
Part (b)). This result nests and explains the---at first sight seemingly contradictory---results by \cite{powers1990limiting} and \cite{amiet2021pure}, which each fall into one of the two identified regimes.
\end{enumerate}

The common framework behind these results is our \cref{thm:master}. It concerns the classification of action profiles of a game  into two groups by any property---in our applications, the indicator that a profile is a pure $\epsilon$-equilibrium. If the classification only exhibits sufficiently weak and local dependence (that is, indicators of action profiles that differ in many coordinates are independent, and those of nearby profiles are only weakly correlated), then 
we show that the asymptotic behavior of the sum of the indicators is then governed by its mean. If the mean converges to a constant then the sum is asymptotically Poisson, if the mean diverges then it is asymptotically Gaussian and positive with probability converging to 1. Each of our results then reduces to computing the expected number of pure $\epsilon$-equilibria while bounding the dependence between nearby profiles under each of our models for generating utilities (i)-(iv), and this is what allows seemingly disparate results to be understood within a common framework. The proof of the \cref{thm:master} combines the Chen--Stein method (cf.\ \citealp*{arratia1989two}) with Stein's method for locally dependent variables (cf.\ \citealp*{demborinott1996}).

\subsection*{Additional related literature}

The study of bounded rationality goes back to the foundational work of \cite{Simon1947}; for surveys, see \cite*{conlisk1996bounded}, \cite*{artinger2022satisficing}, and \cite*{Clippel2024}. Several strands of the literature posit that humans settle on outcomes that are ``approximately optimal'', with  explanations including status quo bias \citep{thaler1980toward} and focusing on differences of first-order importance \citep{gigerenzer1996reasoning}. In a game, this is captured by $\epsilon$-equilibrium, where $\epsilon$ may be interpreted as the level of deviation from optimizing behavior.

Various studies have investigated $\epsilon$-equilibrium. For example, $\epsilon$-equilibrium has been explored to address the discontinuity in equilibria between finitely and infinitely repeated games \citep{radner1980collusive}, to approximate equilibria by studying games with restricted strategy sets \citep{fudenberg1986limit},  to approximate Nash equilibria in perturbed games or in games where agents lack common knowledge or common beliefs \citep*{monderer1989approximating,kajii1998payoff,jackson2012epsilon}, and to investigate their computational properties \citep*{goldberg2014communication,babichenko2017communication}. 
Regarding existence, several articles establish sufficient conditions that guarantee the existence of pure 
$\epsilon$-equilibrium, namely, that each single agent's influence on another agent's utility is limited or that  
the utility of each agent depends only on the
aggregate behavior of the other agents \citep{azrieli2013lipschitz,kalai2004large,carmona2009existence}.  
Our approach is complementary because games that adhere to these sufficient conditions have measure zero in our settings. 
Therefore, our findings neither imply nor are implied by the aforementioned existence results.

To make general statements about the structure of strategic environments without restricting attention to stylized subclasses, a body of literature considers the space of all strategic environments (endowed with an appropriate measure that results from drawing utilities from an underlying distribution). Examples include matching \citep*{ashlagi2017unbalanced}, network games \citep{jackson2007diffusion},  percolation games \citep{Gar23}, extensive form games \citep*{arieli2016random}, and normal form games \citep*{goldberg1968probability,Rin00,alon2021dominance,pradelski2024satisficing}.
We adopt this approach to study the existence of pure $\epsilon$-equilibrium.

Many of the aforementioned works consider large economic environments, that is, with a large number of agents. Large economic environments generally ensure analytical tractability and predictive power. Examples of large economic environments arise in the context of cooperative games \citep{aumann1974values}, auctions \citep*{bodoh2013efficiency} and double auctions \citep*{rustichini1994convergence}, voting \citep{feddersen1997voting},  extensive form games \citep{kalai2004large}, and  market design \citep{azevedo2019strategy}.

\section{Setup}

For any positive integer $x$, write $[x]:= \{1,\dots,x\}$.

\subsection{Model}\label{sec:model}

\begin{definition*}[Grids]
For any integer $n\ge2$, an \emph{$n$-dimensional grid} is a set of the form $\mathbf A = A_1 \times \cdots \times A_n$ where, for each $i \in [n]$, $A_i = [\alpha_i]$ for some integer $\alpha_i \geq 2$. We refer to $A_i$ as the \emph{$i$-th factor} of $\mathbf A$ and write $\alpha_i := |A_i|$ for its cardinality. For any integers $n, M \geq 2$, we write $\mathcal A_M(n)$ for the set of $n$-dimensional grids with $\max_{i\in [n]}\alpha_i \leq M$.
\end{definition*}

\begin{definition*}[Games]
For any integer $n \geq 2$  and an $n$-dimensional grid $\mathbf A$, a \emph{game} 
\[
\left( \, \mathbf A \, , \, \{u_i\}_{i \in [n] } \, \right)
\]
is a tuple where $[n]$ is the agent set, $\mathbf A$ is the set of action profiles (with the $i$th factor $A_i$ being the set of actions of agent $i$), and $u_i: \mathbf A \to \mathbb R$ is agent $i$'s utility.
\end{definition*}

\begin{definition*}[Pure $\epsilon$-equilibrium]
    For any $\epsilon \geq 0$ and game $g=(\mathbf A , \{u_i\}_{i \in [n] })$, an action profile $\mathbf a \in \mathbf A$ is a \emph{pure $\epsilon$-equilibrium} of $g$ if, for each agent $i \in [n]$ and $x\in A_i$, $u_i(\actionvec)\geq u_i(x,\actionvec_{-i})-\epsilon$.\footnote{We write $(x,\actionvec_{-i}) \in \mathbf A$ for an action profile where $x$ is $i$th action and $\actionvec_{-i} $ is the action profile of agents in $[n] \setminus \{i\}$.}
\end{definition*}

\noindent For $\epsilon=0$, pure $\epsilon$-equilibrium reduces to pure Nash equilibrium.

\subsection{The prevalence of pure \texorpdfstring{$\epsilon$}{epsilon}-equilibria}
\label{sec:prevalence}

For a given number of agents $n$, we draw utilities at random from a specified joint distribution and analyze the asymptotic (in $n$) distribution and asymptotic probability of existence of pure $\epsilon$-equilibria.
Since features of the game may vary with $n$, such as the numbers of actions or parameters governing how utilities are drawn, we specify, for each $n$, the set of
configurations that we allow, and we establish our results uniformly over that set.

Formally, for all integers $n,M\ge2$, we consider index sets of the form
\[
\mathcal I(n) = \mathcal A_M(n) \times\mathcal J(n)
\]
consisting of pairs $(\mathbf A,j)$, where the grid $\mathbf A\in\mathcal A_M(n)$ fixes the shape of the game (how many actions each agent has) and $j\in\mathcal J(n)$ supplies whatever further parameters are needed to pin down the law of its utilities. For each $(\mathbf A,j)\in\mathcal I(n)$ we let
\[
\bigl(u_i(\mathbf a)\bigr)_{i\in[n], \mathbf a\in\mathbf A}\sim\mathbb J_{n,(\mathbf A,j)}
\;\text{for some distribution }\mathbb J_{n,(\mathbf A,j)}\text{ on }\mathbb R^{\,n|\mathbf A|},
\]
where the correspondence $(\mathbf A,j)\mapsto\mathbb J_{n,(\mathbf A,j)}$ is specified separately before each of our results. What $\mathcal J(n)$ contains varies accordingly: in the simplest model (\cref{sec:independent_continuous}) $\mathcal J(n)$ is a singleton consisting of the marginal distribution from which utilities are independently drawn,  while elsewhere it records
 the marginal distribution and the strength of correlation across agents (\cref{sec:correlation}), the marginal distribution and interaction graph (\cref{sec:graphical}), or the resolution at which payoffs are measured (\cref{sec:independent_discrete}).

Our objects of interest throughout the paper are the distributions of the number of pure $\epsilon$-equilibria and their corresponding standardized count under $\mathbb J_{n, (\mathbf A,j)}$. Respectively, for each $x \in \mathbb R$, these are defined as: 
\[
\mathbb P_{n, (\mathbf A,j)}^\epsilon(x):= \Pr\big( N^\epsilon(g) \leq x \big) \;\text{ and }\; \tilde{\mathbb P}_{n,(\mathbf A,j)}^\epsilon(x):= \Pr\big( \tilde N^\epsilon(g) \leq  x \big)
\]
where $g = (\mathbf A , \{u_i\}_{i \in [n] })$ is a game whose utilities are drawn according to $\mathbb J_{n,(\mathbf A,j)}$, $N^\epsilon(g)$ is the number of pure $\epsilon$-equilibria in $g$, and $\tilde N^\epsilon(g) $ is the standardized count\footnote{The standardized count, and hence $\tilde{\mathbb P}^{\epsilon}_{n,(\mathbf A,j)}$, is defined whenever $\operatorname{Var}[N^\epsilon(g)]>0$, which holds for all sufficiently large $n$ and every $(\mathbf A,j)\in\mathcal I(n)$ in each of our normal-approximation results.}
\[
\tilde N^\epsilon(g) := \frac{N^\epsilon(g) - \mathbb E[N^\epsilon(g)]}{\sqrt{ \text{Var}[N^\epsilon(g)] }} .
\]
Our results concern asymptotic (in $n$) properties of $\mathbb P_{n, (\mathbf A,j)}^\epsilon$ and $\tilde{\mathbb P}_{n,(\mathbf A,j)}^\epsilon$ that are uniform in $(\mathbf A,j) \in \mathcal I(n)$. To succinctly express our results, we introduce further notation below.

\subsection{Notation}

\begin{definition*}[$\Longrightarrow$ for real numbers]
For real numbers $c_{n,\m}$ defined for every sufficiently large integer $n$ and every $\m \in \mathcal I(n)$, where $\mathcal I(n)$ is a non-empty index set for each such $n$, and for any real number $c$, we write
\[
c_{n,\m} \Longrightarrow  c \qquad \text{if and only if} \qquad \lim_{n \to \infty} \sup_{\m \in \mathcal I(n)} |c_{n,\m} - c  | = 0 .
\]
For divergence, we write $c_{n,\m} \Longrightarrow \infty$ if and only if $\lim_{n\to\infty} \inf_{\m\in\mathcal I(n)} c_{n,\m}=\infty$.
\end{definition*}

\noindent The notation $\Longrightarrow$ is simply shorthand for convergence of $c_{n, \m}$ to $c$ in $n$ that is uniform over $\m \in \mathcal I(n)$. We also extend the use of $\Longrightarrow$ to real-valued functions as follows.

\begin{definition*}[$\Longrightarrow$ for real-valued functions]
For real-valued functions $f_{n,\m}$ defined for every sufficiently large integer $n$ and every $\m \in \mathcal I(n)$, where $\mathcal I(n)$ is a non-empty index set for each such $n$, and for any real-valued function $f$, we write
\[
f_{n,\m} \Longrightarrow  f \qquad \text{if and only if} \qquad \lim_{n \to \infty} \sup_{\m \in \mathcal I(n)} \sup_{x \in \mathbb R}|f_{n,\m}(x) - f(x)  | = 0 .
\]
\end{definition*}
\noindent In other words, $f_{n,\m}\Longrightarrow f$ if $f_{n,\m}$ converges to $f$ as $n\to\infty$, uniformly in $x\in\mathbb R$ and uniformly over $\m\in\mathcal I(n)$.

\begin{definition*}[Gaussian ($\Phi$) and Poisson ($\Pi_\lambda$) distributions]
For any $x \in \mathbb R$ we write $ \Phi(x):=\Pr[\operatorname{Normal}(0,1) \leq x] $ for the cdf of a standard Gaussian random variable. 
For any $x \in \mathbb R$ and any $\lambda \geq  0$, we write $ \Pi_\lambda(x) := \Pr[\mathrm{Poisson}(\lambda) \leq x] $ for the cdf of a Poisson random variable with parameter $\lambda$.
\end{definition*}

\section{Results on the existence and distribution of pure \texorpdfstring{$\epsilon$}{epsilon}-equilibria}\label{sec:main}

In this section we state and discuss our results on the existence and distribution of pure $\epsilon$-equilibria.

\subsection{Independent, continuous case}\label{sec:independent_continuous}

We first study the case where utilities are independent both across agents and across action profiles and payoff ties occur with probability zero.

Fix an integer $M\ge2$ and a continuous distribution $F$ on $\mathbb R$, and for each $n \geq 2$ and $(\mathbf A, F) \in \mathcal I(n) := \mathcal A_M(n) \times \{F\}$, let the $(u_i(\mathbf a))_{i\in[n], \mathbf a\in \mathbf A}$ be iid draws from $F$. Observe that the set $\mathcal J(n)=\{F\}$ is therefore a singleton and the joint utility distribution $\mathbb J_{n,(\mathbf A, F)}$ has a product structure with continuous marginals $F$.

\begin{theorem}[Independent, continuous case]\label{thm:independent_continuous}
Fix $\epsilon \geq 0$, an integer $M \geq 2$, and a continuous distribution $F$ on $\mathbb R$. Suppose that for all $n\geq 2$ and $(\mathbf A, F) \in \mathcal I(n) :=\mathcal A_M(n) \times \{F\}$, the joint utility distribution $\mathbb J_{n, (\mathbf A, F)}$ corresponds to utilities being iid from $F$.
\begin{enumerate}[(a),topsep=0ex,itemsep=0ex,leftmargin=0cm]
\item If $\epsilon = 0$ then 
\[
\mathbb P_{n,(\mathbf A, F)}^0 \Longrightarrow \Pi_1
\]
and, consequently, $\mathbb P_{n,(\mathbf A, F)}^0(0) \Longrightarrow \frac{1}{e}$. That is, when utilities are iid from a continuous distribution $F$, as the number of agents $n$ gets large, the number of pure Nash equilibria converges in distribution to $\operatorname{Poisson}(1)$ and, consequently, the probability that a game admits a pure Nash equilibrium converges to $1-1/e$.

\item If $\epsilon > 0$ then $\mathbb P^{\epsilon}_{n,(\mathbf A, F)}(0) \Longrightarrow 0$ and if, in addition, $\epsilon < \operatorname{diam}(\operatorname{supp} F)$ then
\[
\tilde{\mathbb P}^{\epsilon}_{n,(\mathbf A, F)} \Longrightarrow \Phi .
\]
That is, when utilities are iid from a continuous distribution $F$, as the number of agents $n$ gets large, the probability that a game admits a pure $\epsilon$-equilibrium converges to $1$; and if, moreover, $\epsilon<\operatorname{diam}(\operatorname{supp} F)$ then the standardized number of pure $\epsilon$-equilibria converges in distribution to the standard Gaussian.
\end{enumerate}
\vspace{0.2cm}
\noindent 
In cases (a) and (b), by definition of $\Longrightarrow$, convergence is uniform over numbers of actions bounded by $M$.
\end{theorem}

\cref{thm:independent_continuous} Part (a) recovers the classic result, which is due to \cite{arratia1990poisson} (cf. also \citealp{Rin00}); namely, the distribution of pure Nash equilibria tends to Poisson(1) as the number of agents gets large and, consequently, the probability that a game admits a pure Nash equilibrium converges to $1-1/e\approx 0.63$. 

\cref{thm:independent_continuous} Part (b), which is new, shows that for any fixed $\epsilon>0$ the probability that a game admits a pure $\epsilon$-equilibrium converges to 1 as the number of agents gets large (and the standardized number of such equilibria becomes standard Gaussian).\footnote{If $\epsilon \geq \operatorname{diam}(\operatorname{supp} F)$ then every outcome would be a pure $\epsilon$-equilibrium.} In other words, relaxing the requirement of optimization by allowing agents to be within any $\epsilon >0$ of their respective optima ensures the general existence of stable outcomes.

\subsection{Correlation across agents}\label{sec:correlation}

We next ask whether the asymptotic existence of pure $\epsilon$-equilibria established in \cref{sec:independent_continuous} is robust to correlation between utilities. We first study utilities that are correlated across agents but not across action profiles. 

Fix an integer $M\ge2$ and, for each $n\ge2$, a non-empty set $\mathcal L(n)\subseteq [-1,\infty)$ of \emph{correlation-tuning parameters}, and define the index set
\[
\mathcal I(n):= \mathcal A_M(n) \times \mathcal L(n) .
\]
For each $n \geq 2$ and $(\mathbf A,\lambda)\in\mathcal I(n)$ let the utilities $(u_i(\mathbf a))_{i\in[n],\mathbf a\in\mathbf A}$ be given by
\[
u_i(\mathbf a) := v_i(\mathbf{a}) + \frac{\lambda}{n} \sum_{j\in [n]} v_j(\mathbf{a}) ,
\]
where the values $(v_i(\mathbf a))_{i\in[n], \mathbf a\in\mathbf A}$ are iid uniform on the interval $\Omega:= [-1/2,1/2]$.\footnote{Shifting the interval $\Omega$ by a constant shifts every utility by the same amount and so leaves the number of pure $\epsilon$-equilibria unchanged. Centering the interval at zero is therefore simply a convenience and comes without loss of generality. The assumption that the distribution of the values is uniform is primarily a convenience: our proofs rely on it but we suspect they extend to any continuous distribution $F$ on $\Omega$. The boundedness of the domain is, however, a genuine restriction, and it avoids technical difficulties that would arise relating to tail decay. The specific interval length (of 1) is convenient because the value of $\epsilon$ has a direct relation to the utility scale.} The parameter $\lambda$ tunes the correlation in utilities across agents: a direct computation gives that for any $\mathbf a$ and $i \neq j$,
\[
\vartheta:=\operatorname{Corr}[u_i(\mathbf a),u_j(\mathbf a)]=\frac{(1+\lambda)^{2}-1}{n+(1+\lambda)^{2}-1 },
\]
which ranges over $[-1/(n-1),1)$ as $\lambda$ ranges over $[-1,\infty)$; this is precisely the admissible common correlations for $n$ exchangeable random variables. Correlation is negative when $\lambda <0$ and positive when $\lambda>0$, and $\lambda=0$ recovers the independence model of \cref{sec:independent_continuous} (with the continuous distribution $F$ corresponding to a uniform over $\Omega$). When $\lambda=-1$, every game whose utilities are drawn according to $\mathbb J_{n, (\mathbf A,\lambda)}$ is a zero-sum game. When $\lambda$ grows at a rate of at least $\sqrt{n}$ the correlation parameter is positive, $\vartheta \in (0,1)$.\footnote{Our model does not capture common interest games ($\vartheta=1$), where the existence of pure Nash
equilibria is guaranteed \citep{monderer1996potential} and their number is asymptotically normal \citep*{baldi1989normal}. The latter conclusion can also be obtained from our \cref{thm:master}, combined with a direct lower bound on the variance of the number of pure Nash equilibria.}

We first provide sufficient conditions for the existence of pure $\epsilon$-equilibria.

\begin{theorem}[Correlation across agents, existence]\label{thm:correlation}
Fix $\epsilon \geq 0$ and an integer $M \geq 2$, and suppose that for all $n\geq2$ and $(\mathbf A,\lambda) \in \mathcal I(n) := \mathcal A_M(n) \times \mathcal L(n)$, the joint utility distribution $\mathbb J_{n,(\mathbf A,\lambda)}$ corresponds to utilities being set according to the correlation model with parameter $\lambda$.
\begin{enumerate}[(a),topsep=0ex,itemsep=0ex,leftmargin=0cm]
\item If $\epsilon = 0$ and there exists some constant $c > 0$ such that $\mathcal L(n) \subseteq [c,\infty)$ for all sufficiently large $n$, then
\[
\tilde{\mathbb P}^{0}_{n,(\mathbf A, \lambda)} \Longrightarrow \Phi 
\]
and, moreover, $\mathbb P^0_{n,(\mathbf A,\lambda)}(0) \Longrightarrow 0$. 
That is, when utilities are drawn according to a positive correlation-tuning parameter that is uniformly bounded away from zero, as the number of agents $n$ gets large, the standardized number of pure Nash equilibria converges in distribution to a standard Gaussian and, moreover, the probability that a game admits a pure Nash equilibrium converges to $1$.
\item If $\epsilon > 0$ and there exists some constant $c' > -2\epsilon$ such that $\mathcal L(n) \subseteq [c',\infty)$ for all sufficiently large $n$, then $\mathbb P^{\epsilon}_{n,(\mathbf A,\lambda)}(0) \Longrightarrow 0$ and if, in addition, $\epsilon<1$ then
\[
\tilde{\mathbb P}^{\epsilon}_{n,(\mathbf A, \lambda)} \Longrightarrow \Phi. 
\]
That is, when utilities are drawn according to a correlation-tuning parameter that is bounded below by a constant larger than $-2\epsilon$ (in particular, any non-negative one), as the number of agents $n$ gets large, the probability that a game admits a pure $\epsilon$-equilibrium converges to $1$; and if, moreover, $\epsilon<1$, then the standardized number of pure $\epsilon$-equilibria converges in distribution to a standard Gaussian. 
\end{enumerate}
\vspace{0.2cm}
\noindent 
In cases (a) and (b), by definition of $\Longrightarrow$, convergence is uniform over numbers of actions bounded by $M$ and over correlation-tuning parameters.
\end{theorem}

The condition $\epsilon<1$ in \cref{thm:correlation} Part (b) mirrors the condition on the diameter in \cref{thm:independent_continuous} Part (b). 
We now turn to sufficient conditions for the \emph{non}-existence of pure $\epsilon$-equilibria.

\begin{theorem}[Correlation across agents, non-existence]\label{thm:correlation-nonexistence}
Fix $\epsilon\in[0,\tfrac12)$, an integer $M\ge2$, and a constant $\varsigma\in(0,1)$, such that
\[
\mathcal A_M(n;\epsilon,\varsigma):=\Bigl\{\mathbf A\in\mathcal A_M(n)\ :\
\Bigl(\textstyle\prod_{i\in[n]}\alpha_i\Bigr)^{1/n}
\bigl(\tfrac12+\epsilon\bigr)^{\min_{i\in[n]}\alpha_i-1}\ \le\ \varsigma\Bigr\}
\]
is non-empty for every $n\ge2$. For each $n\ge2$ let $\mathcal I(n):=\mathcal A_M(n;\epsilon,\varsigma)\times\mathcal L(n)$ with $\mathcal L(n):=\{-1\}$, and suppose that for all $n\ge2$ and $(\mathbf A,\lambda)\in\mathcal I(n)$ the joint utility distribution $\mathbb J_{n,(\mathbf A,\lambda)}$ corresponds to utilities being set according to the correlation model with parameter $\lambda=-1$. Then $\mathbb P^{\epsilon}_{n,(\mathbf A,\lambda)}(0)\Longrightarrow1$. That is, in zero-sum games in which the agents have sufficiently many actions, as the number of agents $n$ gets large, the probability that a game admits a pure $\epsilon$-equilibrium converges to $0$. By definition of $\Longrightarrow$, convergence is uniform over $\mathcal I(n)$.
\end{theorem}

\noindent The conditions of \cref{thm:correlation-nonexistence} are, for example, satisfied when every agent has the same number $\alpha$ of actions with $\alpha\bigl(\tfrac12+\epsilon\bigr)^{\alpha-1}\le\varsigma$ (for $\epsilon=0$, one can take $\varsigma=\tfrac34$, which makes the inequality hold for every $\alpha\ge3$). 

\cite{Rin00} studied the existence of pure Nash equilibria under an equivalent correlation model.\footnote{In their model, the marginals are Gaussian rather than uniform. As they mention, the choice of marginal distribution does not alter their results.} 
They derive results for a fixed correlation parameter (where our tuning parameter $\lambda$ grows like $\sqrt{n}$) and for the zero-sum case ($\lambda=-1$). \cref{thm:correlation} Part (a) permits $\mathcal L(n)$ to contain tuning parameters growing at rate $\sqrt n$ thereby recovering the positive correlation case of \cite{Rin00}, but also allows any other rate. 

Turning to negative correlation, \cref{thm:correlation} Part (b) shows that asymptotic existence of pure $\epsilon$-equilibria for $\epsilon>0$ is determined by a tolerance-correlation trade-off: negative correlation pushes towards few or no pure $\epsilon$-equilibria while the $\epsilon$-tolerance pushes towards many. When the $\epsilon$-tolerance exceeds half the magnitude of the tuning parameter existence is ensured. Finally, \cref{thm:correlation-nonexistence} extends \cite{Rin00} by showing that in zero-sum games asymptotic non-existence of pure $\epsilon$-equilibria holds even for small but positive values of $\epsilon$. 
 In fact, the threshold on $\epsilon$ is tight for zero-sum games with sufficiently many actions.\footnote{To see this, note that \cref{thm:correlation} Part (b) with $\mathcal L(n)=\{-1\}$ applies, with $c'=-1$, whenever $-1>-2\epsilon$, that is, whenever $\epsilon>\tfrac12$, and then gives existence of pure $\epsilon$-equilibrium with probability converging to $1$. Conversely, \cref{thm:correlation-nonexistence} gives the opposite conclusion for every $\epsilon<\tfrac12$ when all agents have the same number $\alpha$ of actions, provided $\alpha$ is large enough that $\alpha(\tfrac12+\epsilon)^{\alpha-1}<1$. Hence, the constant threshold on $\epsilon$ is tight.}

\subsection{Correlation across action profiles}\label{sec:graphical}

We next consider the other dimension of dependence: utilities that are correlated across action profiles. To do so in a principled way, we follow \cite{daskalakis2011connectivity} in considering graphical games \citep{kearns2007graphical}. In such games, an agent's utility depends only on her own action and on the actions of her neighbors in an undirected interaction graph. Hence, an agent's utilities are perfectly correlated across action profiles whenever $i$ and her neighbors act identically under $\actionvec$ and $\actionvec'$, because then $u_i(\actionvec)=u_i(\actionvec')$ holds.

Like \cite{daskalakis2011connectivity}, we restrict our analysis to games on expander graphs: an undirected graph $G$ on a vertex set $[n]$ is a $\gamma$-expander  if
$
|\, G(S) \,|\ge\min\{n,\gamma |S|\}
$
for every $S\subseteq [n]$ with $|S|\le\lceil n/\gamma\rceil$, and where $G(S):=\cup_{i\in S}G(i)$ with $G(i)$ denoting the neighborhood of $i$, excluding $i$, in $G$ \citep[cf.][]{hoory2006expander}. Every agent of a $\gamma$-expander has at least $\gamma$ neighbors, and the joint neighborhoods of small sets of agents are relatively large.

For any $c >0$ and integer $n \geq 2$, let $\mathcal G_c(n)$ denote the set of all undirected graphs on the vertex set $[n]$ that are $\gamma$-expanders for some $\gamma \geq c \ln n$.\footnote{Observe that $\mathcal G_c(n)$ is non-empty if and only if $c\ln n\le n-1$, which holds for all sufficiently large $n$.} Fix also a continuous distribution $F$ on $\mathbb R$ and an integer $M \geq 2$, and define the index set\footnote{In the notation of \cref{sec:prevalence}, $\mathcal I(n)=\mathcal A_M(n)\times\mathcal J(n)$ with $\mathcal J(n)=\mathcal G_c(n)\times\{F\}$ itself a product; as is standard, we write elements of $\mathcal I(n)$ as triples $(\mathbf A,G,F)$ rather than as pairs $(\mathbf A,(G,F))$.}
\[
\mathcal I(n):= \mathcal A_M(n) \times \mathcal G_c(n) \times \{F \} .
\]
 For each $n\ge2$ and $(\mathbf A,G,F) \in \mathcal I(n)$, let the joint utility distribution $\mathbb J_{n,(\mathbf A,G,F)}$ correspond to utilities being set according to the following \emph{graphical games model}: for each agent $i$ and each sub-profile $\mathbf x \in \prod_{j \in G(i)  \cup \{i\}} A_j$, draw a value $v_i(\mathbf x)$ from $F$ independently across agents and across sub-profiles, and set $i$'s utility at action profile $\mathbf a \in \mathbf A$ in the resulting graphical game with interaction graph $G$ to be $u_i(\mathbf a):= v_i(\mathbf{a}_{G(i)  \cup \{i\}})$. 
Observe that if $G$ is the complete graph, then the joint utility distribution is the same as the one described in \cref{sec:independent_continuous}, where utilities are iid from a continuous distribution $F$.

\begin{theorem}[Correlation across action profiles]\label{thm:graphical}
Fix $\epsilon \geq 0$, $c > 0$, an integer $M \geq 2$, and a continuous distribution $F$ on $\mathbb R$. For each $n \geq 2$ and $(\mathbf A,G,F) \in \mathcal I(n):= \mathcal A_M(n) \times \mathcal G_c(n) \times \{F \}$, let the joint utility distribution $\mathbb J_{n,(\mathbf A,G,F) }$ correspond to utilities being set according to the graphical games model with parameter $c$.
\begin{enumerate}[(a),topsep=0ex,itemsep=0ex,leftmargin=0cm]
\item If $\epsilon = 0$ and $c > 1/\ln 2$ then 
\[
\mathbb P_{n,(\mathbf A,G,F) }^0 \Longrightarrow \Pi_1
\]
and, consequently, $\mathbb P_{n,(\mathbf A,G,F) }^0(0) \Longrightarrow \frac{1}{e}$. That is, if utilities are set according to the graphical games model with parameter $c> 1/\ln 2$, as the number of agents $n$ gets large, the number of pure Nash equilibria converges in distribution to \textnormal{Poisson(1)} and, consequently, the probability that a game admits a pure Nash equilibrium converges to $1-1/e$.

\item If $\epsilon>0$ and $c > 1/\ln(1/\xi_{F,2}^\epsilon)$ then
\[
\tilde{\mathbb P}^{\epsilon}_{n,(\mathbf A,G,F) } \Longrightarrow \Phi
\]
and, moreover, $\mathbb P^{\epsilon}_{n,(\mathbf A,G,F) }(0) \Longrightarrow 0$. That is, if utilities are set according to the graphical games model with parameter $c> 1/\ln(1/\xi_{F,2}^\epsilon)$, as the number of agents $n$ gets large, the standardized number of pure $\epsilon$-equilibria converges in distribution to the standard Gaussian; and, moreover, the probability that a game admits a pure $\epsilon$-equilibrium converges to 1.
\end{enumerate}
\vspace{0.2cm}
\noindent 
In cases (a) and (b), by definition of $\Longrightarrow$, convergence is uniform over numbers of actions bounded by $M$ and over interaction graphs with logarithmic (at least $c \ln n$) expansion.
\end{theorem}

\begin{remark*}
    The thresholds in parts (a) and (b) are consistent because $\xi_{F,2}^0 = \tfrac12$.
\end{remark*}

\cref{thm:graphical} Part (a) recovers and extends \citet*[Theorem 1.13]{daskalakis2011connectivity}, who establish the $\Poisson(1)$ limit of pure Nash equilibria for binary-action graphical games with expansion $(1+\eta)\log_2 n$ for some $\eta>0$ (which is precisely our condition, $c>1/\ln 2$, since $c\ln n=(c\ln 2)\log_2 n$). Our result, however, applies to games with any (uniformly bounded) number of actions per agent.

\cref{thm:graphical} Part (b) considers the case $\epsilon>0$. In this case, the probability that a game admits a pure $\epsilon$-equilibria converges to 1 as the number of agents gets large (and the standardized number of such equilibria becomes standard Gaussian).

\subsection{Independent, discrete case}\label{sec:independent_discrete}
Having studied departures from independence in \cref{sec:correlation,sec:graphical}, we return to the independence model of \cref{sec:independent_continuous} and consider utility ties. Pure $\epsilon$-equilibria invite an interpretation under which agents optimize exactly, but utilities are measured at a finite resolution. In other words,  utilities within $\epsilon$ of each other are indistinguishable. This raises the question of the extent to which similar results may be obtained under a model in which $\epsilon = 0$ but indifferences arise with positive probability. We therefore now consider a model in which utilities are independent across agents and across action profiles, but indifferences  occur with positive probability.

Fix an integer $M\ge2$ and, for each $n\ge2$, a non-empty set $\mathcal R(n)\subseteq\{2,3,\dots\}$ of payoff resolutions, and consider the index set
\[
\mathcal I(n):= \mathcal A_M(n) \times \mathcal R(n) .
\]
For each $n \geq 2$ and $(\mathbf A ,\rho) \in\mathcal I(n)$ let the utilities $(u_i(\mathbf a))_{i\in[n],\mathbf a\in \mathbf A}$ be iid uniform on $[\rho]$.

\begin{theorem}[Independent, discrete case]\label{thm:independent_discrete}
Fix an integer $M \geq 2$, and suppose that for all $n\geq2$ and $(\mathbf A ,\rho) \in \mathcal I(n):= \mathcal A_M(n) \times \mathcal R(n)$, the joint utility distribution $\mathbb J_{n, (\mathbf A ,\rho)}$ corresponds to utilities being iid uniform from $[\rho]$.
\begin{enumerate}[(a),topsep=0ex,itemsep=0ex,leftmargin=0cm]
\item If $n/\rho \Longrightarrow 0$ then 
\[
\mathbb P^{0}_{n, (\mathbf A ,\rho)} \Longrightarrow \Pi_1 
\]
and, consequently, $\mathbb P^{0}_{n, (\mathbf A ,\rho)}(0) \Longrightarrow \frac{1}{e}$. That is, when utilities are iid uniform on $[\rho]$, as the number of agents $n$ gets large, if  the payoff resolution grows faster than the number of agents, the number of pure Nash equilibria converges in distribution to $\operatorname{Poisson}(1)$ and, consequently, the probability that a game admits a pure Nash equilibrium converges to $1-1/e$.

\item If $n/\rho \Longrightarrow \infty$ then
\[
\tilde{\mathbb P}^{0}_{n, (\mathbf A ,\rho)} \Longrightarrow \Phi
\]
and, moreover, $\mathbb P^{0}_{n, (\mathbf A ,\rho)}(0) \Longrightarrow 0$. That is, when utilities are iid uniform on $[\rho]$, as the number of agents $n$ gets large, if the payoff resolution grows slower than the number of agents, the standardized number of pure Nash equilibria converges in distribution to the standard Gaussian; and, moreover, the probability that a game admits a pure Nash equilibrium converges to 1.
\end{enumerate}
\vspace{0.2cm}
\noindent 
In cases (a) and (b), by definition of $\Longrightarrow$, convergence is uniform over numbers of actions bounded by $M$ and over payoff resolutions.
\end{theorem}

Special cases of our model for generating indifferences were previously studied in the literature. \cref{thm:independent_discrete} unifies and explains the different results that were previously obtained. \citet*{powers1990limiting} considered a model in which payoffs are drawn with replacement from $\{1,\dots,|\mathbf A|=\prod_{i\in [n]}\alpha_i\}$, which implies that the payoff resolution grows exponentially in $n$, and found that pure Nash equilibria converges to Poisson(1). This is nested under Part (a). By contrast, \citet*{amiet2021pure} studied a model in which the payoff resolution remains fixed, and found that the number of pure Nash equilibria grows exponentially and satisfies a central limit theorem. This is nested under Part (b). \cref{thm:independent_discrete} thus locates the transition between these two strands at a payoff resolution of order $n$.

There is, in fact, a specific parametrization of $\epsilon$ that establishes a formal equivalence between the model considered in this section (which is $\epsilon$-free) and the independent-continuous model (with $\epsilon$-dependence) considered in \cref{sec:independent_continuous}. We elaborate on this equivalence in \cref{sec:thm-ind-disc-grid}.

\section{The \texorpdfstring{\cref{thm:master}}{Master Theorem}}\label{sec:master}

Our \cref{thm:master} abstracts away from the language of games (in particular, utilities) and instead studies grids (see \cref{sec:model}), where each point of the grid is endowed with a random indicator. In our applications of the \cref{thm:master} the grid corresponds to the set of action profiles, and the indicator at each profile records whether that profile is a pure $\epsilon$-equilibrium.

\begin{definition*}[Indicator grids]
For any grid $\mathbf A$, we refer to a family of random indicator variables $(I_{\mathbf a})_{\mathbf a \in \mathbf A}$ as an \emph{indicator grid}; each profile $\mathbf a \in \mathbf A$ is associated with a random indicator $I_{\mathbf a} \in \{0,1\}$.
\end{definition*}

We next define well-behaved indicator grids. In such grids, the indicators $(I_{\mathbf a})_{\mathbf a}$ may be correlated, but the correlation strength must be mild and limited to profiles that are close to each other. 

For any two profiles $\mathbf a$ and $\mathbf a'$, let $d_H(\mathbf a,\mathbf a')$ denote the Hamming distance between $\mathbf a$ and $\mathbf a'$; this is the number of entries at which the two profiles differ. Moreover, for any positive integer $r$, let $B_r(\mathbf a):= \{\mathbf a' :d_H(\mathbf a,\mathbf a') \leq r\}$ denote the set of profiles at a Hamming distance at most $r$ from $\mathbf a$.

\begin{definition*}[Well-behaved indicator grids]
For an integer $r \geq 1$ and a real number $\theta \geq 0$, we say that an indicator grid $(I_{\mathbf a})_{\mathbf a\in\mathbf A}$ is \emph{$(r,\theta)$-well-behaved} if:
\begin{enumerate}[(H1),topsep=1ex,itemsep=0ex]
\item\label{hyp:range} \emph{Finite dependence range.} For all $U,U'\subseteq\mathbf A$ such that $d_H(\mathbf a,\mathbf a')> r$ for every $\mathbf a\in U$ and $\mathbf a'\in U'$, the families $(I_{\mathbf a})_{\mathbf a\in U}$ and $(I_{\mathbf a'})_{\mathbf a'\in U'}$ are independent of each other.
\item\label{hyp:homog} \emph{Symmetry.} For every $\mathbf a,\mathbf a'\in\mathbf A$, $\mathbb E[I_{\mathbf a}]=\mathbb E[I_{\mathbf a'}]$.
\item\label{hyp:pair} \emph{Local pair correlation.} For every $\mathbf a\in\mathbf A$,
\[
\sum_{\mathbf a'\in B_{r}(\mathbf a)\setminus\{\mathbf a\}} \mathbb E\bigl[I_{\mathbf a}I_{\mathbf a'}\bigr]\ \le\ \mathbb E[I_{\mathbf a}]\,\theta .
\]
\end{enumerate}
\end{definition*}

\ref{hyp:range} bounds the dependence \emph{range}: indicators of profiles that differ in more than $r$ coordinates are independent, so all dependence is confined to Hamming balls of radius $r$. \ref{hyp:homog} assumes symmetry: every profile's indicator has the same expectation. Finally, \ref{hyp:pair} bounds the \emph{strength} of the dependence that \ref{hyp:range} permits: within a Hamming ball, the sum of probabilities that two profiles are counted simultaneously is bounded by some $\theta$.

\begin{remark}\label{rem:balls}
For any grid $\mathbf A$ and any integer $r \geq 1$, observe that $| B_r(\mathbf a) |$ is independent of $\mathbf a$. Hence we can define 
    \[
    \beta_{\mathbf A}(r) := | B_r(\mathbf a) \,|,
    \]
    which counts the number of profiles at a Hamming distance at most $r$ from any profile.
\end{remark}

Let $\mathcal A(n)$ denote the set of all $n$-dimensional grids.

\begin{mastertheorem}\refstepcounter{mastertheorem}\label{thm:master}
For each integer $n\ge2$ let $\mathcal J(n)$ be an index set and let $\mathcal I(n)\subseteq\mathcal A(n)\times\mathcal J(n)$, and suppose that $\mathcal I(n)$ is non-empty for all sufficiently large $n$. 

For all sufficiently large $n$ and each $\m=(\mathbf A, j) \in\mathcal I(n)$, let $(I_{\mathbf a})_{\mathbf a\in\mathbf A}$ be an $(r_{n,\m},\theta_{n,\m})$-well-behaved indicator grid. 

Define $S_{n,\m}:=\sum_{\mathbf a\in\mathbf A}I_{\mathbf a}$, $\mu_{n,\m}:=\mathbb E[S_{n,\m}]$, $\sigma_{n,\m}:=\sqrt{\operatorname{Var}[S_{n,\m}]}$ and, for each $x\in\mathbb R$, let
\[
\mathbb P_{n,\m}(x):=\Pr\bigl[S_{n,\m}\le x\bigr]
\qquad\text{and}\qquad
\tilde{\mathbb P}_{n,\m}(x):=\Pr\Bigl[\tfrac{S_{n,\m}-\mu_{n,\m}}{\sigma_{n,\m}}\le x\Bigr] \; \; (\text{whenever $\sigma_{n,\m}>0$)}.
\]
Moreover, define $b_{n,\m}:=\beta_{\mathbf A}(r_{n,\m})$ and $p_{n,\m}:=\frac{\mu_{n,\m}}{|\mathbf A|}$. 

Then the following hold:
\begin{enumerate}[(i),topsep=0ex,itemsep=0ex,leftmargin=0cm]
\item\label{thm:master:poisson}
\emph{(Poisson approximation.)} If $\mu_{n,\m}\Longrightarrow c\in[0,\infty)$ and
$\mu_{n,\m}\bigl(b_{n,\m}p_{n,\m}+\theta_{n,\m}\bigr)\Longrightarrow0$, then
$\mathbb P_{n,\m}\Longrightarrow\Pi_{c}$; in particular
$\mathbb P_{n,\m}(0)\Longrightarrow e^{-c}$. 

\item\label{thm:master:clt}
\emph{(Normal approximation.)} If $\mu_{n,\m}\Longrightarrow\infty$ and
\[
\min\left\{\ \frac{\mu_{n,\m}\,b_{n,\m}^{2}}{\sigma_{n,\m}^{3}}\ ,\ \
\mu_{n,\m}\bigl(b_{n,\m}p_{n,\m}+\theta_{n,\m}\bigr)+\theta_{n,\m}\ \right\}\Longrightarrow0,
\]
where we adopt the convention $\mu_{n,\m}b_{n,\m}^2/\sigma_{n,\m}^3:=\infty$ when $\sigma_{n,\m}=0$, then $\sigma_{n,\m}>0$ for all sufficiently large $n$ and all $\m \in\mathcal I(n)$, and $\tilde{\mathbb P}_{n,\m}\Longrightarrow\Phi$.

\item\label{thm:master:existence}
\emph{(Existence.)} If $\mu_{n,\m}\Longrightarrow\infty$ and $\theta_{n,\m}/\mu_{n,\m}\Longrightarrow0$, then $\mathbb P_{n,\m}(0)\Longrightarrow0$.
\end{enumerate}
\end{mastertheorem}

In all the models that we considered in \cref{sec:independent_continuous,sec:independent_discrete,sec:graphical,sec:correlation}, for any $(\mathbf A,j) \in \mathcal I(n) =\mathcal A_M(n)\times\mathcal J(n)$, a game $g$ whose utilities are drawn from $\mathbb J_{n,(\mathbf A,j)}$ induces an indicator grid $(I_{\mathbf a})_{\mathbf a \in \mathbf A}$ where $I_{\mathbf a}$ is the indicator that $\mathbf a$ is a pure $\epsilon$-equilibrium of $g$. Hence $S_{n,(\mathbf A,j)}=N^\epsilon(g)$, and $\mathbb P_{n,(\mathbf A,j)}$ and $\tilde{\mathbb P}_{n,(\mathbf A,j)}$ are precisely our objects of interest, as described in \cref{sec:prevalence}.\footnote{The \cref{thm:master} applies to any property of action profiles that induces a well-behaved indicator grid (such as shown for the case of pure $\epsilon$-equilibrium). For example, the asymptotic existence of satisficing equilibria in random games can also be obtained as an application. Satisficing equilibrium is an ordinal relaxation that requires each agent to play one of her top-ranked actions, irrespective of how much utility she forgoes. While similar in spirit, it is conceptually distinct from pure $\epsilon$-equilibrium \citep{pradelski2024satisficing}.}

The \Cref{thm:master} reduces the proofs of \cref{thm:independent_continuous,thm:independent_discrete,thm:graphical,thm:correlation,thm:correlation-nonexistence} to two steps. In the first step, we show that the induced indicator grid is well-behaved. \ref{hyp:homog} holds immediately for all the models we consider. The dependence range of \ref{hyp:range} is 1 in the models with independence across both agents and across action profiles, 2 in the model with correlation across agents, and it is growing with $n$ in the model with graphical games. We can similarly characterize the strength parameter $\theta$ of \ref{hyp:pair} for each of our models. In the second step, we show that the asymptotic conditions of the \cref{thm:master} hold. The details are provided in \cref{sec:proof-of-theorems}.

The proof of the \cref{thm:master} itself, which is in \cref{sec:proof_T5}, builds on the Chen--Stein theorem \citep[cf.][]{arratia1989two}, which bounds the distance between the distribution of a sum of dependent Bernoulli random variables and that of a Poisson random variable, and on a theorem due to  \cite{demborinott1996}, which bounds the distance between the distribution of a sum of dependent Bernoulli random variables and that of a standard normal variable. The bounds, in each case, are functions of the parameters of well-behaved indicator grids.

\section*{Appendix}

\appendix

\section{Probabilistic preliminaries}\label{app:preliminaries}

The following probabilities, $\xi_{F,\alpha}^\epsilon$ and $\xi_{R,\alpha}$, play an important role in our proofs.

\begin{definition*}[$\xi_{F,\alpha}^\epsilon$ and $\xi_{R,\alpha}$]
For any $\epsilon\geq 0$, any integer $\alpha \geq 2$, and any continuous distribution $F$ on $\mathbb R$, let
\[
\xi_{F,\alpha}^\epsilon:=\Pr\bigl[X_1\ge\max\{X_2,\dots,X_\alpha\}-\epsilon\bigr],
\]
where $X_1,\dots,X_\alpha$ are iid draws from $F$. For integers $\alpha,\rho\ge2$, let
\[
\xi_{\rho,\alpha}:=\Pr\bigl[X_1\ge\max\{X_2,\dots,X_\alpha\}\bigr],
\]
where $X_1,\dots,X_\alpha$ are iid uniform on $[\rho]$.
\end{definition*}

\begin{lemma}\label{lem:xi-cont}
For any integer $\alpha \geq 2$ and continuous distribution $F$ on $\mathbb R$, $\xi_{F,2}^\epsilon \geq \xi_{F,\alpha}^\epsilon > \xi_{F,\alpha}^0 = 1/\alpha$ for any $\epsilon >0$ and if $\epsilon < \operatorname{diam}(\operatorname{supp} F)$ then $\xi_{F,2}^\epsilon<1$.
\end{lemma}

\begin{proof}[Proof of \cref{lem:xi-cont}]
Let $X_1,\dots,X_\alpha$ be iid draws from $F$ and $Y:=\max\{X_2,\dots,X_\alpha\}$. Conditioning on $X_1$ gives $\xi_{F,\alpha}^\epsilon=\int F(x+\epsilon)^{\alpha-1}\,dF(x)$, which is non-increasing in $\alpha$ and, since $F(X_1)$ is uniform on $[0,1]$, equals $1/\alpha$ when $\epsilon=0$. For $\epsilon>0$, $\xi^\epsilon_{F,\alpha}-\xi^0_{F,\alpha}=\Pr[0<Y-X_1\le\epsilon]$, which is positive.\footnote{With
positive probability all $\alpha$ draws lie within $\epsilon/2$ of a given point of the support of $F$,
and conditional on this, since $F$ has no atoms, $X_1$ is not the largest draw with probability
$1-1/\alpha$.} Finally, $\xi^\epsilon_{F,2}=1-\Pr[X_2-X_1>\epsilon]$, and if $\epsilon<\operatorname{diam}(\operatorname{supp}F)$ there are points $a<b$ of the support with $b-a>\epsilon$, so with positive probability $X_1$ lies close enough to $a$ and $X_2$ close enough to $b$ that $X_2-X_1>\epsilon$. Hence $\xi^\epsilon_{F,2}<1$.
\end{proof}

\begin{lemma}\label{lem:xi-disc}
For all integers $\alpha,\rho \geq 2$,
\[
\frac{1}{\alpha}+\frac{1}{2\rho}\ \le\ \xi_{\rho,\alpha}  \leq \min \left\{ \frac{3}{4}, \frac{1}{\alpha}\,\Bigl(1+\frac{1}{2\rho}\Bigr)^{\!\alpha}  \right\}.
\]
\end{lemma}
\begin{proof}[Proof of \cref{lem:xi-disc}]
Observe that
\[
\xi_{\rho,\alpha}:=\Pr\bigl[X_1\ge\max\{X_2,\dots,X_\alpha\}\bigr] =\rho^{-\alpha}\sum_{k=1}^{\rho}k^{\alpha-1},
\]
where $X_1,\dots,X_\alpha$ are iid uniform on $[\rho]$. For the first upper bound, note that $\xi_{\rho,\alpha}$ is non-increasing in $\alpha$ so $\xi_{\rho,\alpha} \leq \xi_{\rho,2} = \rho^{-2}\sum_{k=1}^{\rho}k=\frac{\rho+1}{2\rho} \leq \frac34$. To obtain the other bounds, we set $g(x):=x^{\alpha-1}$, which is convex on $[0,\infty)$ since $\alpha\ge2$, and satisfies $g(0)=0$, and we use the Hermite--Hadamard inequality: for a convex function $g$ and reals $a<b$, $ g\left(\frac{a+b}{2}\right) \leq \frac{1}{b-a}\int_a^b g(x)\,dx \leq \frac{g(a)+g(b)}{2} $. Applying the right-hand inequality on $[k-1,k]$ for each $k\in[\rho]$ and summing,
\[
\frac{\rho^\alpha}{\alpha}=\int_0^{\rho}g(x)\,dx\ \le\ \sum_{k=1}^{\rho}\frac{g(k-1)+g(k)}{2} =\sum_{k=1}^{\rho}g(k)-\frac{g(\rho)-g(0)}{2} =\sum_{k=1}^{\rho}g(k)-\frac{\rho^{\alpha-1}}{2}
\]
Rearranging and dividing by $\rho^{\alpha}$ gives the lower bound. For the remaining upper bound, apply the left-hand Hermite-Hadamard inequality on $[k-\tfrac12,k+\tfrac12]$, whose midpoint is $k$ and whose length is one, gives $g(k)\le\int_{k-1/2}^{k+1/2}g(x)\,dx$ for each $k\in[\rho]$. Summing over $k\in[\rho]$, the intervals tile $[\tfrac12,\rho+\tfrac12]$, so
\[
\sum_{k=1}^{\rho}k^{\alpha-1}\ \le\ \int_{1/2}^{\rho+1/2}x^{\alpha-1}\,dx
=\frac{\bigl(\rho+\tfrac12\bigr)^{\alpha}-\bigl(\tfrac12\bigr)^{\alpha}}{\alpha}
\ \le\ \frac{\bigl(\rho+\tfrac12\bigr)^{\alpha}}{\alpha},
\]
and dividing by $\rho^{\alpha}$ yields $\xi_{\rho,\alpha}\le\tfrac1\alpha\bigl(1+\tfrac1{2\rho}\bigr)^{\alpha}$.
\end{proof}

The rest of \cref{app:preliminaries} collects probabilistic results that are employed in the proofs of our theorems.

\begin{lemma}\label{lem:poisson_be}
There is a universal constant $C_2$ such that for every $\lambda\ge1$ and
$X\sim\operatorname{Poisson}(\lambda)$,
\[
\sup_{x\in\mathbb R} \left| \,\Pr\!\left[\tfrac{X-\lambda}{\sqrt{\lambda}}\leq x\right] -\Phi(x) \, \right| \le \frac{C_2}{\sqrt{\lambda}}.
\]
\end{lemma}
\begin{proof}[Proof of \cref{lem:poisson_be}]
Since the sum of $n$ iid $\operatorname{Poisson}(\theta)$ variables is $\operatorname{Poisson}(n\theta)$, we can write $X=\sum_{i=1}^{\lceil\lambda\rceil}X_i$, where the $(X_i)_i$ are iid $\operatorname{Poisson}(\theta)$ with $\theta:=\lambda/\lceil\lambda\rceil\in[1/2,1]$. Letting $Y \sim \operatorname{Poisson}(\theta)$, we have $\mathbb E[Y]=\operatorname{Var}[Y]=\theta$ and
\[
\mathbb E\bigl[|Y -\theta|^3\bigr] \le \Bigl(\mathbb E\bigl[(Y-\theta)^4\bigr]\Bigr)^{3/4} =\left(\theta+3\theta^2 \right)^{3/4} \le 4^{3/4} \le 3,
\]
where the first step is an application of Lyapunov's inequality, the second step is the Poisson central-moment identity, and the final step follows from $\theta \leq 1$.

The Berry--Esseen theorem for sums of iid random variables therefore gives, with a universal constant $C_0$,
\[
\sup_{x\in\mathbb R} \left| \,\Pr\!\left[\tfrac{X-\lambda}{\sqrt{\lambda}}\leq x\right] -\Phi(x) \, \right| \leq C_0 \,\frac{\lceil\lambda\rceil\cdot\mathbb E\bigl[|Y-\theta|^3\bigr]}{\lambda^{3/2}} \leq C_0 \frac{2 \lambda \cdot 3}{\lambda^{3/2}} = \frac{C_2}{\sqrt{\lambda}},
\]
and thus the claim holds with $C_2=6C_0$.
\end{proof}

\begin{lemma}\label{lem:poisson-tv}
For all $\lambda,\lambda'\geq 0$,
\[ 
\sup_{x\in\mathbb R}\, \Bigl|\,\Pi_\lambda(x)-\Pi_{\lambda'}(x) \,\Bigr|\le|\lambda-\lambda'|.
\]
\end{lemma}

\begin{proof}[Proof of \cref{lem:poisson-tv}]
Assume without loss of generality $\lambda<\lambda'$, and let $X\sim\Poisson(\lambda)$ and $Y\sim\Poisson(\lambda'-\lambda)$ be independent, so that $X+Y\sim\Poisson(\lambda')$. For every $x$ the events $\{X\le x\}$ and $\{X+Y\le x\}$ coincide on $\{Y=0\}$, so $|\Pi_\lambda(x)-\Pi_{\lambda'}(x)|\le\Pr[Y\ge1]=1-e^{-(\lambda'-\lambda)}\le\lambda'-\lambda$.
\end{proof}

\begin{lemma}\label{lem:phi-scale}
For every $\rho>0$ with $|\rho-1|\le \frac12$,
\[
\sup_{x\in\mathbb R} \,\Bigl|\,\Phi(\rho x)-\Phi(x) \,\Bigr|\leq \frac{|\rho-1|}{2}.
\]
\end{lemma}

\begin{proof}[Proof of \cref{lem:phi-scale}]
By the mean value theorem $\Phi(\rho x)-\Phi(x)=\varphi(y)(\rho-1)x$ for some $y$ between $x$ and $\rho x$, where $\varphi$ is the standard normal density. Since $\rho\ge1/2$, $|y|\ge|x|/2$, so $\varphi(y)\le(2\pi)^{-1/2}e^{-x^{2}/8}$ and
\[
\bigl|\Phi(\rho x)-\Phi(x)\bigr|\le\frac{|\rho-1|}{\sqrt{2\pi}}|x|e^{-x^{2}/8}
\le\frac{|\rho-1|}{\sqrt{2\pi}}\sup_{z\ge0}ze^{-z^{2}/8}
=\frac{2e^{-1/2}}{\sqrt{2\pi}}|\rho-1|\le\frac{|\rho-1|}{2}. \qedhere
\]
\end{proof}

\begin{lemma}\label{lem:corr-tails}
Let $X_1,\dots,X_n$ be iid uniform on $[-\tfrac12 , \tfrac12]$ and $\bar X:=\tfrac1n\sum_{i=1}^n X_i$. Then $\Pr[\bar X\ge x]\ge\tfrac12(1-2x)^{n}$ for every $x\in[0,\tfrac12)$.
\end{lemma}
\begin{proof}[Proof of \cref{lem:corr-tails}]
Let $E:=\bigcap_i\{X_i\ge2x-\tfrac12\}$. This is an event of probability $(1-2x)^{n}$. Conditionally on $E$ the variables $X_i-(2x-\tfrac12)$ are iid uniform on $[0,1-2x]$, so $\bar X-(2x-\tfrac12)$ has a distribution symmetric about $\tfrac12-x$ and therefore
$\Pr[\bar X\geq x \mid E]=\Pr[\bar X-(2x-\tfrac12)\ge\tfrac12-x\mid E]\ge\tfrac12$.
\end{proof}

\begin{lemma}\label{lem:sep}
Let $Z$ be a random element of a set $\mathcal Z$, let $f:\mathcal Z\to\mathbb R$ be bounded, and consider $A,B\subseteq\mathcal Z$ with $\Pr[Z\in A]\ge p$ and $\Pr[Z\in B]\ge p$. Suppose that for some $k\in\mathbb R$ we have $f(z)\le k$ for every $z\in A$ and
$f(z)\ge k+1$ for every $z\in B$. Then $\operatorname{Var}[f(Z)]\ge p/4$.
\end{lemma}

\begin{proof}[Proof of \cref{lem:sep}]
Write $m:=\mathbb E[f(Z)]$ and suppose first that $m\le k+\tfrac12$. For every $z\in B$ we have $f(z)-m\ge k+1-m\ge\tfrac12$, both sides being non-negative, so $(f(z)-m)^{2}\ge\tfrac14$; hence $\operatorname{Var}[f(Z)]=\mathbb E\bigl[(f(Z)-m)^{2}\bigr]
\geq \mathbb E\bigl[\mathbbm 1\{Z\in B\}\,(f(Z)-m)^{2}\bigr] \geq \tfrac14\Pr[Z\in B]\ \geq p/4$. If instead $m>k+\tfrac12$, then for every $z\in A$ we have $m-f(z)\ge m-k>\tfrac12$, and the same computation with $A$ in place of $B$ gives $\operatorname{Var}[f(Z)]\ge\tfrac14\Pr[Z\in A]\ge p/4$.
\end{proof}

\begin{theorem}[Chen--Stein; cf.\ \citealp*{arratia1989two}, Theorem 1] \label{thm:chenstein-orig}
Let $\mathcal J$ be a finite index set and $(I_\alpha)_{\alpha\in\mathcal J}$
random variables with values in $\{0,1\}$ and and $\Pr[I_\alpha=1]>0$ for each $\alpha$. Write $W:=\sum_{\alpha}I_\alpha$. For
each $\alpha\in\mathcal J$, let $\mathcal B(\alpha)\subseteq\mathcal J$ with
$\alpha\in \mathcal B(\alpha)$ be such that $I_\alpha$ is independent of the family
$(I_\beta)_{\beta\notin \mathcal B(\alpha)}$. Then
\[
\sup_X\bigl|\,\Pr[W\in X]-\Pr[\operatorname{Poisson}(\mathbb E[W])\in X]\,\bigr|\le c_1+c_2,
\]
where
\[
c_1=\sum_{\alpha\in\mathcal J}\sum_{\beta\in \mathcal B(\alpha)} \mathbb E[I_\alpha]\,\mathbb E[I_\beta],
\qquad
c_2=\sum_{\alpha\in\mathcal J}\sum_{\beta\in \mathcal B(\alpha)\setminus\{\alpha\}} \mathbb E[I_\alpha I_\beta].
\]
\end{theorem}

\begin{theorem}[\citealp{demborinott1996}, Theorem 4.2]\label{thm:demborinott}
Let $\mathcal J$ be a finite index set and $(I_\alpha)_{\alpha\in\mathcal J}$
random variables satisfying $|I_\alpha-\mathbb E [I_\alpha]|\le c$ almost
surely, and write $W:=\sum_{\alpha}I_\alpha$, $\sigma^{2}:=\operatorname{Var}[W]>0$, and $\bar\mu:=|\mathcal J|^{-1}\sum_{\alpha}\mathbb E[\, |I_\alpha-\mathbb E [I_\alpha] | \, ]$. For each $\alpha\in\mathcal J$ let $\mathcal B(\alpha)\subseteq\mathcal J$, where $\alpha'\in\mathcal B(\alpha)$ if and only if $\alpha\in\mathcal B(\alpha')$, be such that $(I_\alpha,I_{\alpha'})$ is independent of $(I_\beta)_{\beta\notin\mathcal B(\alpha)\cup\mathcal B(\alpha')}$ for all $\alpha,\alpha'\in\mathcal J$, and set $d:=\max_{\alpha}|\mathcal B(\alpha)|$. Then 
\[
\sup_{x\in\mathbb R}\,\Bigl|\,\Pr\Bigl[\tfrac{W-\mathbb E[W]}{\sigma}\leq x \Bigr]-\Phi(x) \,\Bigr| \leq \frac{7\,|\mathcal J|\,\bar\mu\,(d c)^{2}}{\sigma^{3}} .
\]
\end{theorem}

\section{Properties of (stochastic) grids}

\begin{lemma}\label{lem:ball-power}
    Consider an $n$-dimensional grid $\mathbf A = A_1 \times \cdot \cdot \times A_n$ and recall that $\alpha_i:=|A_i|$ for each $i \in [n]$. Then, for every positive integer $r$,
    \[
    \beta_{\mathbf A}(r) := \sum_{s=0}^r  \sum_{\substack{X\subseteq[n]\\ |X|=s}}  \prod_{i\in X}(\alpha_i-1) .
    \]
    And if there is some $M \geq 2$ such that $\alpha_i \leq M$ for each $i \in [n]$ then $\beta_{\mathbf A}(r) \leq \left( en M/r \right)^r$ and $\beta_{\mathbf A}(r)-\beta_{\mathbf A}(r-1)\leq (Mn)^r/r!$. Finally, for every positive integer $k$, $\beta_{\mathbf A}(kr)\le\beta_{\mathbf A}(r)^{k}$.
\end{lemma}
\begin{proof}[Proof of \cref{lem:ball-power}]
    For any $\mathbf a \in \mathbf A$, a profile at Hamming distance $r$ from $\mathbf a$ is obtained by choosing the set $X$ of $r$ coordinates in which it differs from $\mathbf a$ and, for each $i\in X$, one of the $\alpha_i-1$ actions distinct from $a_i$; which gives our formula for $\beta_{\mathbf A}(r)$. For the first upper bound, observe that
    \[
    \beta_{\mathbf A}(r)=\sum_{s=0}^r  \sum_{\substack{X\subseteq[n]\\ |X|=s}}  \prod_{i\in X}(\alpha_i-1) \leq M^r \sum_{s=0}^r \binom{n}{s} \leq \left( \frac{en M}{r} \right)^r ,
    \]
    where the last step follows from a standard combinatorial bound:
    $\sum_{s\le r}\binom{n}{s}\le(en/r)^{r}$.

    For the second upper bound, observe that
    \[
    \beta_{\mathbf A}(r)-\beta_{\mathbf A}(r-1) = \sum_{\substack{X\subseteq[n]\\ |X|=r}}  \prod_{i\in X}(\alpha_i-1) \leq M^r \binom{n}{r} \leq \frac{(nM)^r}{r!},
    \]
    where the final step uses $\binom{n}{r} \leq \frac{n^r}{r!}$.

    For the third upper bound, we first show that $\beta_{\mathbf A}$ is sub-multiplicative: $\beta_{\mathbf A}(s+s')\le\beta_{\mathbf A}(s)\,\beta_{\mathbf A}(s')$ for all integers $s,s'\ge1$. Fix $\mathbf a\in\mathbf A$ and let $\mathbf c\in B_{s+s'}(\mathbf a)$. Let $D\subseteq[n]$ be the set of coordinates at which $\mathbf c$ and $\mathbf a$ differ, so that $|D|\le s+s'$, choose $S\subseteq D$ with $|S|=\min\{s,|D|\}$ and let $\mathbf b$ agree with $\mathbf c$ on $S$ and with $\mathbf a$ on $[n]\setminus S$. Then $\mathbf a$ and $\mathbf b$ differ exactly on $S$, so $d_H(\mathbf a,\mathbf b)=|S|\le s$, and $\mathbf b$ and $\mathbf c$ differ exactly on $D\setminus S$, so $d_H(\mathbf b,\mathbf c)=|D\setminus S|\le s'$. Hence
    \[
    B_{s+s'}(\mathbf a)\ \subseteq\ \bigcup_{\mathbf b\in B_s(\mathbf a)}B_{s'}(\mathbf b),
    \]
    and bounding the size of the union by the sum of the sizes gives, by \cref{rem:balls},
    \[
    \beta_{\mathbf A}(s+s')\ \le\ \sum_{\mathbf b\in B_s(\mathbf a)}\bigl|B_{s'}(\mathbf b)\bigr|
    \ =\ \beta_{\mathbf A}(s)\,\beta_{\mathbf A}(s').
    \]
    Taking $s=r$ and $s'=(k-1)r$, an induction on $k$ yields $\beta_{\mathbf A}(kr)\le\beta_{\mathbf A}(r)^{k}$.
\end{proof}

\begin{lemma}\label{lem:var}
Consider a $(r,\theta)$-well-behaved indicator grid $(I_{\mathbf a})_{\mathbf a\in\mathbf A}$. Let $S:=\sum_{\mathbf a\in\mathbf A}I_{\mathbf a}$, $\mu:=\mathbb E[S]$, $p:=\mu/|\mathbf A|$, and $b:=\beta_{\mathbf A}(r)$. Then
\[
\mu\bigl(1-b\,p\bigr) \le \operatorname{Var}[S] \le \min\bigl\{\,\mu(1+\theta),\ b\mu\,\bigr\} .
\]
\end{lemma}

\begin{proof}
By \ref{hyp:homog} we have $p=\mathbb E[I_{\mathbf a}]$, so that $\operatorname{Var}[I_{\mathbf a}]=p(1-p)$ and $\mu=|\mathbf A|p$. By \ref{hyp:range}, $\operatorname{Cov}[I_{\mathbf a},I_{\mathbf a'}]=0$ whenever $d_H(\mathbf a,\mathbf a')\ge r+1$, so
\begin{equation}\label{eq:var-decomp}
\operatorname{Var}[S]=\mu(1-p) +\sum_{\mathbf a\in\mathbf A} \sum_{\mathbf a'\in B_r(\mathbf a)\setminus\{\mathbf a\}} \operatorname{Cov}[I_{\mathbf a},I_{\mathbf a'}],
\end{equation}
where each inner sum has exactly $b-1$ terms.

First upper bound: since $\operatorname{Cov}[I_{\mathbf a},I_{\mathbf a'}]\le
\mathbb E[I_{\mathbf a}I_{\mathbf a'}]$, \ref{hyp:pair} bounds each inner sum by $p\,\theta$, so the double sum is at most $|\mathbf A|p\,\theta=\mu\theta$, and \eqref{eq:var-decomp} gives $\operatorname{Var}[S]\le\mu(1+\theta)$. Second upper bound: by Cauchy--Schwarz, $\operatorname{Cov}[I_{\mathbf a},I_{\mathbf a'}]\le \sqrt{\operatorname{Var}[I_{\mathbf a}]\operatorname{Var}[I_{\mathbf a'}]}=p(1-p)$, so the double sum is at most $\mu(b-1)(1-p)$ and \eqref{eq:var-decomp} gives $\operatorname{Var}[S]\le\mu(1-p)\,b \le b\mu$. Lower bound: since $I_{\mathbf a}I_{\mathbf a'}\ge0$, we have $\operatorname{Cov}[I_{\mathbf a},I_{\mathbf a'}]\ge-p^{2}$, so the double sum is at least $-\mu(b-1)p$ and \eqref{eq:var-decomp} gives
\[
\operatorname{Var}[S]\ge\mu\bigl[1-p-(b-1)p\bigr] =\mu\bigl(1-bp\bigr).\qedhere
\]
\end{proof}

\section{\texorpdfstring{Proof of the \cref{thm:master} and of \cref{cor:master-clt-suff}}{Proof of the Master Theorem and its corollary}}\label{sec:proof_T5}

Our \cref{thm:master} builds on the Chen--Stein theorem (\cref{thm:chenstein-orig}) and a Stein-type theorem due to Dembo--Rinott (\cref{thm:demborinott}). Below, we state each of these results in a manner that is specialized to well-behaved indicator grids (and, in each case, we show how they follow from their original statements as given in \cref{app:preliminaries}). The statements below are exact inequalities holding for every grid (no index set and no asymptotics are involved). We then obtain the \cref{thm:master} by applying Chen--Stein and Dembo--Rinott for well-behaved indicator grids and considering the large-$n$ behavior of our quantities of interest.

\begin{theorem}[Chen--Stein theorem for well-behaved indicator grids]
\label{thm:chenstein-grid}
Consider a $(r,\theta)$-well-behaved indicator grid $(I_{\mathbf a})_{\mathbf a\in\mathbf A}$. Let $S:=\sum_{\mathbf a\in\mathbf A}I_{\mathbf a}$, $\mu:=\mathbb E[S]$, $p:=\mu/|\mathbf A|$, $b:=\beta_{\mathbf A}(r)$, and let $\mathbb P$ denote the cdf of $S$. Then
\begin{equation*}
  \sup_{x\in\mathbb R}\,\bigl|\,\mathbb P(x)-\Pi_{\mu}(x)\,\bigr|\ \le\ \mu\,\bigl(b\,p+\theta\bigr).
\end{equation*}
\end{theorem}

\begin{proof}[Proof of \cref{thm:chenstein-grid}]
    We apply the original Chen--Stein theorem (\cref{thm:chenstein-orig}) with $\mathcal J = \mathbf A $, $W = S$, and $\mathcal B(\mathbf a) = B_{r}(\mathbf a)$. The required independence of $I_{\mathbf a}$ from $(I_{\mathbf b})_{\mathbf b\notin B_{r}(\mathbf a)}$ holds by \ref{hyp:range}, applied with $U=\{\mathbf a\}$ and $U'=\{\mathbf b:d_H(\mathbf a,\mathbf b)\ge r+1\}$.

    It now remains for us to work out $c_1$ and $c_2$, and to upper-bound them. We start with $c_1$. By \ref{hyp:homog}, for each $\mathbf a \in \mathbf A $, we have $p= \mathbb E[ I_{\mathbf a}]$.\footnote{If $p=0$ then $S=0$ almost surely and $\Pi_0$ is the cdf of the point mass at $0$, so the bound holds trivially. We can therefore take $p>0$.} Hence
    \begin{align*}
    c_1 = \sum_{\mathbf a\in\mathbf A }\sum_{\mathbf a'\in B_{r}(\mathbf a)} \mathbb E[I_{\mathbf a}]\,\mathbb E[I_{\mathbf a'}] &=|\mathbf A |\, b\, p^2 =\mu\, b\, p.
    \end{align*}
    We now turn to $c_2$. By \ref{hyp:pair}, we have 
    \begin{align*}
    c_2&=\sum_{\mathbf a\in \mathbf A }\sum_{\mathbf a'\in B_{r}(\mathbf a)\setminus\{\mathbf a\}} \mathbb E[I_{\mathbf a}I_{\mathbf a'}] 
    \leq |\mathbf A | \, p \, \theta = \mu \,\theta. \qedhere
    \end{align*}
\end{proof}

\begin{theorem}[Dembo--Rinott theorem for well-behaved indicator grids]
\label{thm:demborinott-grid}
Consider a $(r,\theta)$-well-behaved indicator grid $(I_{\mathbf a})_{\mathbf a\in\mathbf A}$ with $\sigma:=\sqrt{\operatorname{Var}[S]}>0$, where $S:=\sum_{\mathbf a\in\mathbf A}I_{\mathbf a}$. Let $\mu:=\mathbb E[S]$, $b:=\beta_{\mathbf A}(r)$, and let $\tilde{\mathbb P}$ denote the cdf of $(S-\mu)/\sigma$. Then
\begin{equation*}
  \sup_{x\in\mathbb R}\,\bigl|\,\tilde{\mathbb P}(x)-\Phi(x)\,\bigr| \ \le\ 14\,\frac{\mu}{\sigma^{3}}\,b^{2} .
\end{equation*}
\end{theorem}

\begin{proof}[Proof of \cref{thm:demborinott-grid}]
We apply the Dembo--Rinott theorem (\cref{thm:demborinott}) with $\mathcal J = \mathbf A$, $W=S$, and $\mathcal B(\mathbf a) = B_r(\mathbf a)$. The sets $\mathcal B(\mathbf a)$ are symmetric, since $\mathbf a'\in B_r(\mathbf a)$ if and only if $\mathbf a\in B_r(\mathbf a')$, and $d=\max_{\mathbf a}|B_r(\mathbf a)|=\beta_{\mathbf A}(r)=b$. The independence requirement is \ref{hyp:range} applied to the families $U=\{\mathbf a,\mathbf a'\}$ and $U'=\mathbf A\setminus\bigl(B_r(\mathbf a)\cup B_r(\mathbf a')\bigr)$, every member of which is at Hamming distance at least $r+1$ from both $\mathbf a$ and $\mathbf a'$. Since $I_{\mathbf a}\in\{0,1\}$ we may take $c=1$ in \cref{thm:demborinott}, and by \ref{hyp:homog} we can write $p = \mathbb E[I_{\mathbf a}]$, so $\bar\mu=\mathbb E[\,|I_{\mathbf a}-p|\,]=2p(1-p)\le2p$ and $|\mathcal J|\bar\mu\le2\mu$. Substituting gives the bound.
\end{proof}

\begin{remark*}
\Cref{thm:demborinott-grid} does not require the indicator grid to satisfy
\ref{hyp:pair}.
\end{remark*}

The three parts of our \cref{thm:master} follow from the Chen--Stein and the Dembo--Rinott theorems for well-behaved indicator grids (\cref{thm:chenstein-grid,thm:demborinott-grid}), as well as from the approximation results of \cref{lem:poisson_be,lem:poisson-tv,lem:phi-scale}.

\begin{proof}[Proof of the \cref{thm:master}]
Throughout we abbreviate $b:=b_{n,\m}$, $\theta:=\theta_{n,\m}$, $\mu:=\mu_{n,\m}$, $p:=p_{n,\m}$, $S:=S_{n,\m}$ and $\sigma:=\sigma_{n,\m}=\sqrt{\operatorname{Var}[S]}$. Fix $\delta\in(0,\tfrac12]$; by definition of $\Longrightarrow$, it suffices in each part to exhibit some $\bar n$ such that the relevant quantity is at most a fixed multiple of $\delta$ for every $n\ge\bar n$ and every $\m\in\mathcal I(n)$.

\medskip\noindent\emph{Part \ref{thm:master:existence}.} Since $\mu\Longrightarrow\infty$ and $\theta/\mu\Longrightarrow0$, choose $\bar n$ such that $\mu\ge1/\delta$ and $\theta/\mu\le\delta$ for all $n\ge\bar n$ and all $\m\in\mathcal I(n)$. As $S\ge0$, we have $\mathbb P_{n,\m}(0)=\Pr[S=0]$, so Chebyshev's inequality and the first upper bound of \cref{lem:var} give
\[
\mathbb P_{n,\m}(0) \leq \Pr\bigl[|S-\mu|\ge\mu\bigr] \leq \frac{\operatorname{Var}[S]}{\mu^{2}} \leq \frac{\mu(1+\theta)}{\mu^{2}} = \frac{1}{\mu}+\frac{\theta}{\mu} \leq 2\delta .
\]

\medskip\noindent\emph{Part \ref{thm:master:poisson}.} By the triangle inequality, \cref{thm:chenstein-grid} and \cref{lem:poisson-tv},
\[
\sup_{x\in\mathbb R}\, \bigl|\,\mathbb P_{n,\m}(x)-\Pi_{\lambda}(x)\,\bigr|  \leq \sup_{x\in\mathbb R}\, \bigl|\,\mathbb P_{n,\m}(x)-\Pi_{\mu}(x)\,\bigr|+\sup_{x\in\mathbb R}\, \bigl|\,\Pi_{\mu}(x)-\Pi_{\lambda}(x)\,\bigr| \leq \mu\bigl(b\,p+\theta\bigr)+\bigl|\mu-\lambda\bigr| .
\]
The first term tends to zero by hypothesis and the second because $\mu\Longrightarrow\lambda$, so $\mathbb P_{n,\m}\Longrightarrow\Pi_{\lambda}$. Evaluating at $x=0$ gives $\mathbb P_{n,\m}(0)\Longrightarrow\Pi_{\lambda}(0)=e^{-\lambda}$.

\medskip\noindent\emph{Part \ref{thm:master:clt}.} Choose $\bar n$ such that, for all $n\ge\bar n$ and all $\m\in\mathcal I(n)$,
\[
\min\Bigl\{\tfrac{\mu\,b^{2}}{\sigma^{3}}\ ,\ \mu(b\,p+\theta)+\theta\Bigr\}\le\delta,
\qquad
\mu\ge1/\delta^{2},
\]
which is possible by the two hypotheses. Fix such $n$ and $\m$, and distinguish two cases according to which term attains the minimum.

\medskip
\noindent Case 1: $\mu\,b^{2}/\sigma^{3}\le\delta$. This case applies only if $\sigma>0$ (if $\sigma=0$ then we take $\mu\,b^{2}/\sigma^{3}$ to be infinite), so $\tilde{\mathbb P}_{n,\m}$ is well defined, and \cref{thm:demborinott-grid} gives directly
\[
\sup_{x\in\mathbb R}\,\bigl|\,\tilde{\mathbb P}_{n,\m}(x)-\Phi(x)\,\bigr|
\ \leq\ 14\,\frac{\mu}{\sigma^{3}}\,b^{2}\ \leq\ 14\,\delta .
\]

\medskip
\noindent Case 2: $\mu(b\,p+\theta)+\theta\le\delta$. Then $b\,p\le\delta/\mu\le\delta^{3}$ and $\theta\le\delta$, so, since $\delta\le\tfrac12$, the lower bound of \cref{lem:var} gives
\begin{equation*}
\sigma^{2}\ \geq\ \mu\bigl(1-b\,p\bigr)\ \geq\ \frac{\mu}{2}\ >\ 0 ,
\end{equation*}
so $\sigma>0$ and $\tilde{\mathbb P}_{n,\m}$ is well defined here too. Writing $\rho:=\sigma/\sqrt\mu$, the two bounds of \cref{lem:var} give $1-b\,p\le\sigma^{2}/\mu\le1+\theta$, and therefore
\begin{equation*}
|\rho-1|=\frac{|\rho^{2}-1|}{\rho+1} \leq \Bigl|\frac{\sigma^{2}}{\mu}-1\Bigr| \leq \max\{b\,p,\ \theta\} \leq \delta.
\end{equation*}Let $X\sim\operatorname{Poisson}(\mu)$ and fix $x\in\mathbb R$. By the triangle inequality, we can write
\[
\bigl|\,\tilde{\mathbb P}_{n,\m}(x)-\Phi(x)\,\bigr| \leq \underbrace{\Bigl|\,\tilde{\mathbb P}_{n,\m}(x)-\Pr\bigl[\tfrac{X-\mu}{\sigma}\le x\bigr]\,\Bigr|}_{\text{(a)}} + \underbrace{\Bigl| \,\Pr\bigl[\tfrac{X-\mu}{\sigma}\le x\bigr]-\Phi(\rho x) \,\Bigr|}_{\text{(b)}} + \underbrace{\bigl|\,\Phi(\rho x)-\Phi(x)\,\bigr|}_{\text{(c)}}
\]
We now bound (a)-(c). We start with (a). Since $\bigl\{(S-\mu)/\sigma\le x\bigr\}=\bigl\{S\le\mu+x\sigma\bigr\}$, \cref{thm:chenstein-grid} gives
\[
\Bigl|\,\tilde{\mathbb P}_{n,\m}(x)-\Pr\bigl[\tfrac{X-\mu}{\sigma}\le x\bigr]\,\Bigr|
=\bigl|\,\mathbb P_{n,\m}(\mu+x\sigma)-\Pi_{\mu}(\mu+x\sigma)\,\bigr|
\ \le\ \mu\bigl(b\,p+\theta\bigr)\ \le\ \delta .
\]
For (b), the definition of $\rho$ combined with \cref{lem:poisson_be}, which is applicable since $\mu\ge1/\delta^{2}\ge1$, give
\[
\Bigl| \,\Pr\bigl[\tfrac{X-\mu}{\sigma}\le x\bigr]-\Phi(\rho x) \,\Bigr| = \Bigl| \,\Pr\bigl[\tfrac{X-\mu}{\sqrt{\mu}}\le \rho x\bigr]-\Phi(\rho x) \,\Bigr|
\ \le\ \frac{C_2}{\sqrt{\mu}}\ \le\ C_2\,\delta .
\]
Finally, for (c), \cref{lem:phi-scale} gives $\bigl|\Phi(\rho x)-\Phi(x)\bigr|\le|\rho-1|/2\le\delta/2$. We therefore conclude that
\[
\sup_{x\in\mathbb R}\,\bigl|\,\tilde{\mathbb P}_{n,\m}(x)-\Phi(x)\,\bigr| \leq \Bigl(\tfrac32+C_2\Bigr)\delta .
\]
In Cases 1 and 2, the bound is a fixed multiple of $\delta$, uniformly over
$\m\in\mathcal I(n)$, so $\tilde{\mathbb P}_{n,\m}\Longrightarrow\Phi$.
\end{proof}

The following corollary provides sufficient condition for the hypotheses of the \cref{thm:master} Part \ref{thm:master:clt} that is simpler to verify.

\begin{corollary}\label{cor:master-clt-suff}
The hypotheses of the \cref{thm:master} Part \ref{thm:master:clt} are satisfied if the following hold: $\mu_{n,\m}\Longrightarrow\infty$, $b_{n,\m}\,p_{n,\m}\Longrightarrow0$ and 
\[
\min\left\{\ \frac{b_{n,\m}^{2}}{\sqrt{\mu_{n,\m}}}\ ,\ \
    \mu_{n,\m}\bigl(b_{n,\m}p_{n,\m}+\theta_{n,\m}\bigr)+\theta_{n,\m}\ \right\} \Longrightarrow 0.
\]
\end{corollary}

\begin{proof}[Proof of \cref{cor:master-clt-suff}]
Fix $\delta\in(0,\tfrac12]$ and take $\bar n$ so that $b_{n,\m}p_{n,\m}\le\tfrac12$ for all $n\ge\bar n$ and all $\m\in\mathcal I(n)$. Then \cref{lem:var} gives $\sigma_{n,\m}^{2}\ge\mu_{n,\m}(1-b_{n,\m}p_{n,\m})\ge\mu_{n,\m}/2>0$,
so if $\frac{b_{n,\m}^{2}}{\sqrt{\mu_{n,\m}}} \Longrightarrow 0$ then $\frac{\mu_{n,\m}\,b_{n,\m}^{2}}{\sigma_{n,\m}^{3}}\ \le\ 2^{3/2}\,\frac{b_{n,\m}^{2}}{\sqrt{\mu_{n,\m}}}\ \Longrightarrow\ 0$. Hence $\min \left\{ \frac{\mu_{n,\m}\,b_{n,\m}^{2}}{\sigma_{n,\m}^{3}}, \mu_{n,\m}\bigl(b_{n,\m}p_{n,\m}+\theta_{n,\m}\bigr)+\theta_{n,\m} \right\} \Longrightarrow 0$ thereby satisfying the hypotheses of the \cref{thm:master} Part \ref{thm:master:clt}.
\end{proof}

\section{\texorpdfstring{Proofs of \cref{thm:independent_continuous,thm:independent_discrete,thm:graphical,thm:correlation,thm:correlation-nonexistence}}{Proofs of the main results}}\label{sec:proof-of-theorems}

\subsection{\texorpdfstring{Proof of \cref{thm:independent_continuous}}{Proof for the independent, continuous case}}

\begin{proposition}[Independent, continuous case]\label{prop:grid-cont}
Fix $\epsilon\geq 0$, an integer $M\ge2$, and a continuous distribution $F$ on $\mathbb R$. For each $n\ge2$ and each $\m = (\mathbf A, F)\in\mathcal I(n)=\mathcal A_M(n) \times \{F\}$ let $I_{\mathbf a}$ be the indicator that $\mathbf a$ is a pure $\epsilon$-equilibrium of $g=(\mathbf A, \{u_i\}_{i \in [n]})$, where $g$ has utilities drawn from $\mathbb J_{n,\m}$ as described in \cref{sec:independent_continuous}. Then we have:
\begin{enumerate}[(a),topsep=0ex,itemsep=0ex,leftmargin=0cm]
\item $(I_{\mathbf a})_{\mathbf a\in\mathbf A}$ is a $(r_{n,\m},\theta_{n,\m})$-well-behaved indicator grid with
\[
r_{n,\m}=1 \text{ and }  \theta_{n,\m}:=n M^2\,\prod_{i\in[n]}\xi_{F,\alpha_i}^\epsilon .
\]
Moreover, letting $S_{n,\m}:=\sum_{\mathbf a\in\mathbf A}I_{\mathbf a}$, $\mu_{n,\m}:=\mathbb E[S_{n,\m}]$, and $p_{n,\m}:=\frac{\mu_{n,\m}}{|\mathbf A|}$,  we have 
\[
\mu_{n,\m}=\prod_{i\in[n]}\alpha_i \,\xi_{F,\alpha_i}^\epsilon , \text{ and } b_{n,\m}:= \beta_{\mathbf A}(r_{n,\m}) \leq nM.
\]
\item If $\epsilon=0$ then the hypotheses of the \cref{thm:master} Part \ref{thm:master:poisson} are satisfied. If $\epsilon >0$ then the hypotheses of the  \cref{thm:master} Part \ref{thm:master:existence} are satisfied; and if, moreover, $\epsilon < \operatorname{diam}(\operatorname{supp} F)$ then the hypotheses of the \cref{thm:master} Part \ref{thm:master:clt} are also satisfied.
\end{enumerate}
\end{proposition}
\noindent \cref{thm:independent_continuous} follows directly from \cref{prop:grid-cont} and the \cref{thm:master}.\hfill$\Box$

As can be seen from the structure of the statement of \cref{prop:grid-cont}, the argument splits into two parts. In part (a) we establish the well-behavedness parameters of the indicator grid induced by the utility distribution $\mathbb J_{n,\m}$. In particular, for this model, the dependence range $r_{n,\m}=1$ and the local pair-correlation $\theta_{n,\m}$ is of order at most $n$. Once all the relevant quantities, including $\mu_{n,\m}$ and $b_{n,\m}$ are derived, the proof of part (b) reduces to taking limits in $n$.

\begin{proof}[Proof of \cref{prop:grid-cont}]
For any $\mathbf a \in \mathbf A \in \mathcal A_M(n)$,
\[
I_{\mathbf a} = \mathbbm 1 \left[ \bigcap_{i \in [n]} \left\{ u_i(\mathbf a)\geq \max_{x \in A_i}u_i(x,\mathbf a_{-i})-\epsilon \right\} \right] .
\]
Since $I_{\mathbf a}$ depends only on the utilities $u_i(\mathbf b)$ with $i\in[n]$ and $\mathbf b\in L^i_{\mathbf a}:=\{(x,\mathbf a_{-i}):x\in A_i\}$, and no such pair $(i,\mathbf b)$ is shared by two profiles at Hamming distance at least $2$, the two families in \ref{hyp:range} with $r=1$ are functions of disjoint sets of independent utilities, and hence are independent. So \ref{hyp:range} holds with $r_{n,\m}=1$. Observe, moreover, that
\[
b_{n,\m}=\beta_{\mathbf A}(1) = 1 + \sum_{i \in [n]} (\alpha_i - 1) \leq 1 + n (M-1) \leq nM .
\]
Since utilities are independent across agents, for any $\mathbf a$, we have that
\[
\mathbb E[I_{\mathbf a}] = \prod_{i\in [n]} \Pr\left[ \, u_i(\mathbf a)\geq \max_{x \in A_i}u_i(x,\mathbf a_{-i})-\epsilon \, \right] = \prod_{i\in [n]} \xi_{F,\alpha_i}^\epsilon =: p_{n,\m} ,
\]
which therefore satisfies \ref{hyp:homog}. Hence $\mu_{n,\m} = |\mathbf A| p_{n,\m} = \prod_{i\in [n]} \alpha_i \xi_{F,\alpha_i}^\epsilon$.

Finally, fix $\mathbf a$ and $\mathbf a'\in B_1(\mathbf a)\setminus\{\mathbf a\}$, and let $i$ be the unique agent with $a_i\neq a'_i$, so that $\mathbf a$ and $\mathbf a'$ both lie on the line $L^i_{\mathbf a}$. Let $E_i$ be the event that both $u_i(\mathbf a)$ and $u_i(\mathbf a')$ are at least $\max_{\mathbf b\in L^i_{\mathbf a}}u_i(\mathbf b)-\epsilon$, and write $\nu^\epsilon_{F,\alpha_i}:=\Pr[E_i]$, which depends only on $F$, $\alpha_i$ and $\epsilon$. Then $I_{\mathbf a}I_{\mathbf a'}=1$ if and only if $E_i$ occurs and, for every $j\neq i$, agent $j$'s $\epsilon$-equilibrium condition $u_j(\mathbf c)\ge\max_{\mathbf b\in L^j_{\mathbf c}}u_j(\mathbf b)-\epsilon$ holds at both $\mathbf c=\mathbf a$ and $\mathbf c=\mathbf a'$. Since all of these events are independent,
\[
\mathbb E\bigl[I_{\mathbf a}I_{\mathbf a'}\bigr]
=\nu^\epsilon_{F,\alpha_i}\prod_{j\in[n]\setminus\{i\}}\bigl(\xi^\epsilon_{F,\alpha_j}\bigr)^2
\le\xi^\epsilon_{F,\alpha_i}\prod_{j\in[n]\setminus\{i\}}\bigl(\xi^\epsilon_{F,\alpha_j}\bigr)^2
=\frac{p_{n,\m}^2}{\xi^\epsilon_{F,\alpha_i}}
\le\alpha_i\,p_{n,\m}^2\le M\,p_{n,\m}^2 ,
\]
where the first inequality holds because $E_i$ is contained in the event $\{u_i(\mathbf a)\ge\max_{\mathbf b\in L^i_{\mathbf a}}u_i(\mathbf b)-\epsilon\}$, whose probability is $\xi^\epsilon_{F,\alpha_i}$, and the second uses $\xi^\epsilon_{F,\alpha_i}\ge1/\alpha_i$ (\cref{lem:xi-cont}). Hence,
\[
\sum_{\mathbf a'\in B_{r_{n,\m}}(\mathbf a)\setminus\{\mathbf a\}} \mathbb E\bigl[I_{\mathbf a}I_{\mathbf a'}\bigr] \leq nM^2 p_{n,\m}^2,
\]
which satisfies \ref{hyp:pair} with $\theta_{n,\m}= nM^2 p_{n,\m}$, and completes the proof of Part (a).

We now turn to Part (b).  Suppose that $\epsilon=0$. By \cref{lem:xi-cont} for any $i$, $\xi_{F,\alpha_i}^0 = 1/\alpha_i$, hence $\mu_{n,\m}=1$. Therefore,
\[
\mu_{n,\m}\bigl(b_{n,\m}p_{n,\m}+\theta_{n,\m}\bigr)\leq p_{n,\m} (nM + nM^2) \leq 2^{-n} (nM + nM^2) \Longrightarrow0,
\]
so the hypotheses of the \cref{thm:master} Part \ref{thm:master:poisson} are satisfied. Suppose instead that $\epsilon>0$. By \cref{lem:xi-cont}, $\tau := \min_{2 \leq \alpha \leq M} \alpha \xi_{F,\alpha}^\epsilon > 1$ whenever $\epsilon >0$, and if $\epsilon < \operatorname{diam}(\operatorname{supp} F)$ then $\xi_{F,2}^\epsilon<1$. It follows that if $\epsilon >0$ then $\mu_{n,\m} \geq \tau^n \Longrightarrow\infty$ and that $\theta_{n,\m}/\mu_{n,\m} = \frac{nM^2}{|\mathbf A|} \leq \frac{nM^2}{2^n}\Longrightarrow0$, thereby satisfying the hypotheses of the \cref{thm:master} Part \ref{thm:master:existence}. If, moreover, $\epsilon < \operatorname{diam}(\operatorname{supp} F)$ then $b_{n,\m}p_{n,\m} \leq nM (\xi_{F,2}^\epsilon)^n \Longrightarrow0$ and $\frac{b_{n,\m}^{2}}{\sqrt{\mu_{n,\m}}} \leq \frac{(Mn)^2}{\tau^{n/2}} \Longrightarrow 0$, thereby satisfying the hypotheses of the \cref{thm:master} Part \ref{thm:master:clt} (by \cref{cor:master-clt-suff}).
\end{proof}

\subsection{\texorpdfstring{Proof of \cref{thm:correlation}}{Proof for correlation across agents: existence}}

\begin{proposition}[Correlation across agents, existence]\label{prop:grid-corr}
Fix $\epsilon \geq 0$, and an integer $M \geq 2$. For each $n\ge2$ and each $\m = (\mathbf A, \lambda)\in\mathcal I(n) = \mathcal A_M(n) \times \mathcal L(n) $, let $I_{\mathbf a}$  be the indicator that $\mathbf a$ is a pure $\epsilon$-equilibrium of $g=(\mathbf A, \{u_i\}_{i \in [n]})$, where $g$ has utilities drawn from $\mathbb J_{n,\m}$ as described in \cref{sec:correlation}. Then we have:
\begin{enumerate}[(a),topsep=0ex,itemsep=0ex,leftmargin=0cm]
\item $(I_{\mathbf a})_{\mathbf a\in\mathbf A}$ is a $(r_{n,\m},\theta_{n,\m})$-well-behaved indicator grid with
\[
r_{n,\m}=2 \text{ and }  \theta_{n,\m}:= (nM)^2 .
\]
Moreover, letting $S_{n,\m}:=\sum_{\mathbf a\in\mathbf A}I_{\mathbf a}$, $\mu_{n,\m}:=\mathbb E[S_{n,\m}]$, and $p_{n,\m}:=\frac{\mu_{n,\m}}{|\mathbf A|}$, we have $b_{n,\m}:= \beta_{\mathbf A}(r_{n,\m}) \leq  (nM)^2$ and
\[
\mu_{n,\m}=\Bigl(\prod_{i\in[n]}  \alpha_i\Bigr) \cdot \int_{\Omega^{n}}\ \prod_{i\in[n]}\Lambda_{n,\m}\bigl(x_i+\lambda \tfrac1n\textstyle\sum_{j \in [n]}x_j +\epsilon\bigr)^{\alpha_i-1}\,dx_1\cdots dx_n,
\]
where $\Lambda_{n,\m}(x):=\Pr[u_i(\mathbf a) \leq x]$ for each $x \in \mathbb R$ (by exchangeability, $\Lambda_{n,\m}$ does not depend on $i$ or $\mathbf a$).
\item Suppose that there is a constant $\lambda_\star>-2\epsilon$ such that
$\mathcal L(n)\subseteq[\lambda_\star,\infty)$ for all sufficiently large $n$. Then the hypotheses
of the \cref{thm:master} Part \ref{thm:master:existence} are satisfied.
\item If, in addition, $\epsilon<1$, then the hypotheses of the \cref{thm:master}
Part \ref{thm:master:clt} are also satisfied.
\end{enumerate}
\end{proposition}
\noindent\cref{thm:correlation} follows directly from \cref{prop:grid-corr} and the \cref{thm:master}.\hfill$\Box$

In this case, the dependence range $r_{n,\m}$ is constant. The expression for the mean $\mu_{n,\m}$ is considerably more complex than in the other models. This has the consequence that establishing parts (b) and, in particular part (c), is more involved and the proof of \cref{prop:grid-corr} requires additional machinery and several lemmas.

Before turning to the proof of \cref{prop:grid-corr}, we state three lemmas, whose hypotheses distinguish two ranges of the correlation parameter: \cref{lem:grid-corr-mu-bounds} lower bounds $\mu_{n,\m}$ for every $\m \in \mathcal I(n)$, while \cref{lem:grid-corr-p-bound} upper bounds $p_{n,\m}$ for those $\m=(\mathbf A,\lambda)$ with $\lambda \le 50M$, and \cref{lem:var-lower-bound} lower bounds $\operatorname{Var}[S_{n,\m}]$ for those with $\lambda > 50M$.

\begin{lemma}\label{lem:grid-corr-mu-bounds}
Suppose that there is a constant $\lambda_\star>-2\epsilon$ such that $\mathcal L(n)\subseteq[\lambda_\star,\infty)$ for all sufficiently large $n$. Let $c_0:=\tfrac1{32}\bigl(\tfrac34\bigr)^{M-1}$, set
\[
s:=\begin{cases}
\min\bigl\{\epsilon+\tfrac{\lambda_\star c_0}{2},\ \tfrac14\bigr\} & \text{if }\lambda_\star>0\\
\min\bigl\{\epsilon+\tfrac{\lambda_\star}{2},\ \tfrac14\bigr\} & \text{if }\lambda_\star\le0
\end{cases}
\]
which is positive in either case, and let $\tau:=1+\tfrac s4>1$. Then there is an integer $N$ such that $\mu_{n,\m}\ge\tau^{n}$ for all $n\ge N$ and all $\m\in\mathcal I(n)$.
\end{lemma}

\begin{lemma}\label{lem:grid-corr-p-bound}
Suppose $\epsilon<1$ and set
\[
\bar q:=\max\Bigl\{1-\tfrac{(1-\epsilon)^{2}}{32},\ e^{-(1-\epsilon)^{2}/(2(50M)^{2})}\Bigr\}\ < 1 .
\]
Then there is an integer $N$ such that $p_{n,\m}\leq 3 \bar q^{n}$ for all $n\geq N$ and all $\m=(\mathbf A,\lambda)\in\mathcal I(n)$ with $\lambda\le 50M$.
\end{lemma}

\begin{lemma}\label{lem:var-lower-bound}
Suppose $\epsilon<1$. Then there is an integer $N$ such that
\[
\operatorname{Var}[S_{n,\m}] \geq \frac{|\mathbf A|}{(nM)^3}\, e^{-n\tfrac{3}{10}\ln 2}
\]
for all $n \geq N$ and all $\m=(\mathbf A,\lambda)\in\mathcal I(n)$ with $\lambda > 50M$.
\end{lemma}

\begin{proof}[Proof of \cref{prop:grid-corr}]
For any $\mathbf a \in \mathbf A$, writing $u_i(\mathbf a) = v_i(\mathbf{a}) + \frac{\lambda}{n} \sum_{j\in [n]} v_j(\mathbf{a}) $, we have
\[
I_{\mathbf a} = \mathbbm 1 \left[ \bigcap_{i \in [n]} \left\{ v_i(\mathbf{a}) + \tfrac{\lambda}{n} \sum_{j\in [n]} v_j(\mathbf{a})  \geq \max_{x \in A_i}u_i(x,\mathbf a_{-i})-\epsilon \right\} \right] .
\]
Since $I_{\mathbf a}$ depends only on the values $v_j(\mathbf b)$ with $j\in[n]$ and $\mathbf b\in B_1(\mathbf a)$, and $B_1(\mathbf a)\cap B_1(\mathbf a')=\emptyset$ whenever $d_H(\mathbf a,\mathbf a')\ge3$, the two families in \ref{hyp:range} with $r=2$ are functions of disjoint sets of independent values, and hence are independent. So \ref{hyp:range} holds with $r_{n,\m}=2$. By \cref{lem:ball-power}, $b_{n,\m}:=\beta_{\mathbf A}(2) \leq \beta_{\mathbf A}(1)^2 = (1 + \sum_{i\in [n]}(\alpha_i - 1))^2 \leq (nM)^2$. By construction, $I_{\mathbf a}$ and $I_{\mathbf a'}$ have the same distribution, so \ref{hyp:homog} holds, and we can let $p_{n,\m}:= \mathbb E[I_{\mathbf a}]$. Since $I_{\mathbf a}I_{\mathbf a'} \leq I_{\mathbf a}$, we have
\[
\sum_{\mathbf a' \in B_2(\mathbf a) \setminus \{\mathbf a\}} \mathbb E[I_{\mathbf a} I_{\mathbf a'}] \leq (\beta_{\mathbf A}(2) -1)\, p_{n,\m} \leq (nM)^2 \,p_{n,\m},
\]
so \ref{hyp:pair} holds with $\theta_{n,\m} = (nM)^2$. Finally, to work out $\mu_{n,\m}=  p_{n,\m} \, |\mathbf A|  = p_{n,\m} \cdot \prod_{i \in [n]} \alpha_i$, it remains for us to derive an explicit expression for $p_{n,\m}$:
 \begin{align*}
 p_{n,\m}&=  \int_{\Omega^n}  \prod_{i \in [n]} \Pr \left[ \max_{a_i'\neq a_i} u_i(a_i',\mathbf{a}_{-i})  \leq x_i + \tfrac{\lambda}{n}\sum_{j\in [n]} x_j + \epsilon  \; \bigg| \; \bigcap_{i=1}^n \{ v_i(\mathbf{a})=x_i\} \right] d x_1 \cdots dx_n  \\
&= \int_{\Omega^n}  \prod_{i\in [n]} \Lambda_{n,\m}\bigl(x_i+\lambda\tfrac1n\textstyle\sum_{j \in [n]}x_j +\epsilon\bigr)^{\alpha_i-1}\,dx_1\cdots dx_n
 \end{align*}   
 The first equality follows from independence across action profiles and conditioning on the values $v_1(\mathbf a),\ldots,v_n(\mathbf a)$. The second equality follows from the fact that the utilities have the same distribution $\Lambda_{n,\m}$ across action profiles, and takes the distribution $\Lambda_{n,\m}(\cdot)^{\alpha_i -1}$ of the maximum of $\alpha_i - 1$ utilities for each $i$. This completes the proof of Part (a).

We now turn to Part (b), and assume that there is a constant $\lambda_\star>-2\epsilon$ such that
$\mathcal L(n)\subseteq[\lambda_\star,\infty)$ for all sufficiently large $n$. By \cref{lem:grid-corr-mu-bounds} there is a constant $\tau>1$ with $\mu_{n,\m}\ge\tau^{\,n}$ for all large $n$ and all $\m\in\mathcal I(n)$. Hence $\mu_{n,\m}\Longrightarrow\infty$ and, by Part (a),
\[
\frac{\theta_{n,\m}}{\mu_{n,\m}}\ \le\ \frac{(nM)^{2}}{\tau^{\,n}}\ \Longrightarrow\ 0 ,
\]
so the hypotheses of the \cref{thm:master} Part \ref{thm:master:existence} are satisfied.

Finally, we turn to Part (c). In addition to the hypothesis of Part (b), we assume that $\epsilon<1$. By \cref{lem:grid-corr-mu-bounds} we still have $\mu_{n,\m}\ge\tau^{\,n}$ for all large $n$ and all $\m\in\mathcal I(n)$, so $\mu_{n,\m}\Longrightarrow\infty$. We bound $\mu_{n,\m}b_{n,\m}^{2}/\sigma_{n,\m}^{3}$ separately on two ranges of the correlation parameter. First, consider $\m=(\mathbf A,\lambda)\in\mathcal I(n)$ with $\lambda\le 50M$. By \cref{lem:grid-corr-p-bound} we have $p_{n,\m}\le3\bar q^{\,n}$ with $\bar q<1$ for all large $n$, so $b_{n,\m}p_{n,\m}\le3(nM)^{2}\bar q^{\,n}\le\tfrac12$ for all large $n$. The variance lower bound of \cref{lem:var} then gives $\sigma_{n,\m}^{2}\ge\mu_{n,\m}\bigl(1-b_{n,\m}p_{n,\m}\bigr)\ge\mu_{n,\m}/2$, so
\[
\frac{\mu_{n,\m}\,b_{n,\m}^{2}}{\sigma_{n,\m}^{3}}
\ \le\ \frac{\mu_{n,\m}\,b_{n,\m}^{2}}{\bigl(\mu_{n,\m}/2\bigr)^{3/2}}
\ =\ \frac{2^{3/2}\,b_{n,\m}^{2}}{\sqrt{\mu_{n,\m}}}
\ \le\ 2^{3/2}(nM)^{4}\,\tau^{-n/2} .
\]
Second, consider $\m=(\mathbf A,\lambda)\in\mathcal I(n)$ with $\lambda>50M$. By \cref{lem:var-lower-bound}, for all large $n$ we have $\sigma_{n,\m}^{2}\ge\frac{|\mathbf A|}{(nM)^{3}}e^{-n\frac{3}{10}\ln2}$. Using $\mu_{n,\m}\le|\mathbf A|$, $b_{n,\m}\le(nM)^{2}$, and $|\mathbf A|\ge2^{\,n}$ (as $\alpha_i\ge2$ for every $i$),
\[
\frac{\mu_{n,\m}\,b_{n,\m}^{2}}{\sigma_{n,\m}^{3}}
\ \le\ \frac{|\mathbf A|\,(nM)^{4}}
{\Bigl(\tfrac{|\mathbf A|}{(nM)^{3}}e^{-n\frac{3}{10}\ln 2}\Bigr)^{3/2}}
\ =\ \frac{(nM)^{17/2}\,e^{\frac{9}{20}n\ln 2}}{|\mathbf A|^{1/2}}
\ \le\ (nM)^{17/2}\,e^{-\frac{n}{20}\ln 2} .
\]
Every $\m\in\mathcal I(n)$ falls into one of these two ranges, so for all large $n$ and all $\m\in\mathcal I(n)$,
\[
\frac{\mu_{n,\m}\,b_{n,\m}^{2}}{\sigma_{n,\m}^{3}}
\ \le\ \max\Bigl\{\,2^{3/2}(nM)^{4}\tau^{-n/2},\ (nM)^{17/2}e^{-\frac{n}{20}\ln 2}\,\Bigr\}
\ \Longrightarrow\ 0 ,
\]
and since the quantity appearing in the hypothesis of the \cref{thm:master} Part \ref{thm:master:clt} is a minimum with $\mu_{n,\m}b_{n,\m}^{2}/\sigma_{n,\m}^{3}$ as one of its arguments, that hypothesis is satisfied. This completes the proof.
\end{proof}

\subsubsection{\texorpdfstring{Proofs of \cref{lem:grid-corr-mu-bounds,lem:grid-corr-p-bound,lem:var-lower-bound}}{Proofs of the lemmas}}

Let $\Lambda(x):=\min\{1,\max\{x+\tfrac12,0\}\}$ denote the cdf of a uniform on $\Omega=[-\tfrac12,\tfrac12]$. We start by recording facts about $\Lambda_{n,\m}$ and $\Lambda$ in \cref{lem:grid-corr-Lambdas} below.

\begin{lemma}\label{lem:grid-corr-Lambdas}
The following hold:
\begin{enumerate}[(i),topsep=0ex,itemsep=0ex]
\item\label{it:sandwich} For every $t\in\mathbb R$, every $n\ge2$, and every $\m=(\mathbf A,\lambda)\in\mathcal I(n)$,
\[
\Lambda\left(t-|\lambda|\sqrt{\tfrac{\ln n}{n}}\right)  -\tfrac{2}{n^{2}} \leq \Lambda_{n,\m}(t) \leq \Lambda\left(t+|\lambda|\sqrt{\tfrac{\ln n}{n}}\right) + \tfrac{2}{n^{2}}.
\]
\item\label{it:mass} For every integer $\alpha\ge2$ and every $s\in[0,1]$,
\[
\alpha\int_\Omega\Lambda(x+s)^{\alpha-1}dx=1+\alpha s-s^{\alpha} \geq 1+s, \text{ and } \int_\Omega\Lambda(x+s)dx=1-\tfrac{(1-s)^{2}}{2}.
\]
\item\label{it:mean} For every integer $\alpha\in\{2,\dots,M\}$ and every $s\in[0,\tfrac14]$,
\[
m_\alpha(s):=\frac{\int_\Omega x\, \Lambda(x+s)^{\alpha-1}dx}{\int_\Omega\Lambda(x+s)^{\alpha-1}dx} \geq c_0:=\tfrac1{32}\bigl(\tfrac34\bigr)^{M-1} >0.
\]
\item\label{it:prob} Let $\bar x = \tfrac1n \sum_{j \in [n]}x_j$ and $c_0:=\tfrac1{32}\bigl(\tfrac34\bigr)^{M-1}$. For every $\m=(\mathbf A,\lambda)\in\mathcal I(n)$, $s\in [0,\tfrac14]$ and all $n \geq 2/c_0^2$,
\[
 \frac{\int_{\{x \in \Omega^n :\, \bar x \geq c_0/2\}} \prod_{i\in [n]}\Lambda(x_i+s)^{\alpha_i-1}\,dx}{\int_{\Omega^n} \prod_{i\in [n]} \Lambda(x_i+s)^{\alpha_i-1}\,dx} \geq \frac{1}{2}.
\]
\end{enumerate}
\end{lemma}

\begin{proof}[Proof of \cref{lem:grid-corr-Lambdas}]
\emph{(i).} Put $\eta:=|\lambda|\sqrt{\ln n/n}$ and $D:=\bigl\{|\bar v(\mathbf a)|\le\sqrt{\ln n/n}\bigr\}$, where $\bar v(\mathbf a):=\tfrac1n\sum_{j\in[n]}v_j(\mathbf a)$, so that $|\lambda\,\bar v(\mathbf a)|\le\eta$ on $D$. Since $u_i(\mathbf a)=v_i(\mathbf a)+\lambda\bar v(\mathbf a)$, on $D$ we have $v_i(\mathbf a)-\eta\le u_i(\mathbf a)\le v_i(\mathbf a)+\eta$. Hence $\{v_i(\mathbf a)\le t-\eta\}\cap D\subseteq\{u_i(\mathbf a)\le t\}\subseteq\{v_i(\mathbf a)\le t+\eta\}\cup D^{c}$. Using $\Pr[A\cap B] \geq \Pr[A]-\Pr[B^c]$ and $\Pr[A\cup B] \leq \Pr[A]+\Pr[B]$,
\[
\Lambda(t-\eta)-\Pr[D^{c}]\ \le\ \Lambda_{n,\m}(t)\ \le\ \Lambda(t+\eta)+\Pr[D^{c}] .
\]
The values $v_j(\mathbf a)$ are iid on $\Omega$ with mean $0$ and range $1$, so Hoeffding's inequality gives $\Pr[D^{c}]=\Pr\bigl[|\bar v(\mathbf a)|>\sqrt{\ln n/n}\bigr]\le2e^{-2n(\ln n/n)}=2n^{-2}$. 

\emph{(ii).} Substituting $u=x+s+\tfrac12$,
\[
\alpha\int_\Omega\Lambda(x+s)^{\alpha-1}dx=\alpha\Bigl[\int_s^{1}u^{\alpha-1}du+s\Bigr]=1+\alpha s-s^{\alpha},
\]
which is at least $1+s$ because $\alpha s\geq 2s$ and $s^{\alpha}\leq s$; setting $\alpha=2$ and dividing by $2$ gives $\tfrac{1+2s-s^{2}}{2}=1-\tfrac{(1-s)^{2}}{2}$.

\emph{(iii).} Write $w(x):=\Lambda(x+s)^{\alpha-1}$, which is non-decreasing with values in $[0,1]$; in particular $\int_\Omega w(x) dx\le1$, so it suffices to bound $\int_\Omega x\,w(x)\,dx$ from below. We have:
\begin{align*}
    \int_\Omega x\,w(x)dx=\int_0^{1/2}x\bigl[w(x)-w(-x)\bigr]dx &\geq \int_{1/4}^{1/2}x\bigl[w(x)-w(-x)\bigr]dx 
    \geq \tfrac1{16}\bigl[w(\tfrac14)-w(-\tfrac14)\bigr]
\end{align*}
where the first step applies the change of variables $x\mapsto-x$ on the negative
half of $\Omega$, the first inequality follows because the integrand is non-negative, and the second inequality uses $x\ge\tfrac14$, $w(x)\ge w(\tfrac14)$ and $w(-x)\le w(-\tfrac14)$ for $x\in[\tfrac14,\tfrac12]$.

Now let $s\in[0,\tfrac14]$, set $a:=\tfrac34+s$ and $b:=\tfrac14+s$, and observe that $b\le\tfrac12a$. Hence, with $k:=\alpha-1\ge1$ we have:
\[
w(\tfrac14)-w(-\tfrac14)=a^{k}-b^{k}=a^{k}\Bigl[1-\bigl(\tfrac ba\bigr)^{k}\Bigr]\ \ge\ \tfrac12\,a^{k}\ \ge\ \tfrac12\bigl(\tfrac34\bigr)^{M-1}.
\]
Therefore $m_\alpha(s)\geq \tfrac1{32}\bigl(\tfrac34\bigr)^{M-1}>0$.

\emph{(iv).} Let $A:=\{x \in \Omega^n : \bar x \geq c_0/2\}$. Observe that
\[
 \frac{\int_{\{x \in \Omega^n :\, \bar x \geq c_0/2\}}\prod_{i\in [n]} \Lambda(x_i+s)^{\alpha_i-1}\,dx}{\int_{\Omega^n}\prod_{i\in [n]} \Lambda(x_i+s)^{\alpha_i-1}\,dx}  = \Pr[(X_1,\dots,X_n) \in A] = \Pr[\bar X \geq c_0/2],
\]
where $\bar X = \tfrac1n \sum_{j \in [n]} X_j$ where the $X_j$s are independent; and for each $i$, $X_i$ is drawn from probability measure on $\Omega$ with density $\Lambda(x+s)^{\alpha_i-1} / \int_{\Omega} \Lambda(y+s)^{\alpha_i-1} dy$. By part (iii), we have that $\mathbb E[X_i] = m_{\alpha_i}(s) \geq c_0$, hence $\mathbb E[\bar X] \geq c_0$. Moreover, each $X_i$ has unit range, $\operatorname{Var}[X_i] \leq \tfrac14$ and therefore $ \operatorname{Var}[\bar X] \leq \tfrac{1}{4n}$. By Chebyshev's inequality,
\[
\Pr[\bar X<c_0/2]\leq \Pr[|\bar X-\mathbb E\bar X|>c_0/2]\leq \tfrac{1/(4n)}{(c_0/2)^{2}}=\tfrac1{nc_0^{2}} .
\]
Hence $\Pr[\bar X \geq c_0/2] \geq 1- \tfrac1{nc_0^{2}} \geq \tfrac12$ for $n\geq 2/c_0^2$.
\end{proof}

\begin{proof}[Proof of \cref{lem:grid-corr-mu-bounds}]
Throughout, write $\eta_n(\lambda):=|\lambda|\sqrt{\ln n/n}$ and $\bar x:=\tfrac1n\sum_{i\in[n]}x_i$ for $x\in\Omega^{n}$.  We split the proof by showing, in turn, that there is an integer $N$ such that, for all $n\ge N$ and all $\m\in\mathcal I(n)$:
\begin{enumerate}[(i),topsep=0ex,itemsep=0ex]
\item if $\eta_n(\lambda)\leq s/2$, then $\mu_{n,\m}\geq \bigl(1+\tfrac s4\bigr)^{n}$;
\item if $\eta_n(\lambda)\geq s/2$, then $\mu_{n,\m}\ge\bigl(1 + \tfrac s4\bigr)^{n}$.
\end{enumerate}

\bigskip
\noindent Case (i): $\eta_n(\lambda)\leq s/2$. Assume $\eta_n(\lambda)\le s/2$ and set $s':=s-\eta_n(\lambda)$, so that $s'\in[\tfrac s2,\tfrac14]$; in particular $s'>0$ and $s'$ lies in the range for which \cref{lem:grid-corr-Lambdas}\ref{it:mean} applies. Define
\[
A:=\begin{cases}
\bigl\{x\in\Omega^{n}: \bar x\geq \tfrac {c_0}{2}\bigr\} & \text{if }\lambda>0,\\
\Omega^{n} & \text{if }\lambda\le0.
\end{cases}
\]
Let $x\in A$ and $i\in[n]$. Then
\begin{align*}
    \Lambda_{n,\m}\bigl(x_i+\lambda\bar x+\epsilon\bigr)\ \ge\ \Lambda\bigl(x_i+\lambda\bar x+\epsilon-\eta_n(\lambda)\bigr)-\tfrac{2}{n^{2}}\ \ge\ \Lambda\bigl(x_i+s'\bigr)-\tfrac{2}{n^{2}} ,
\end{align*}
where the first inequality follows from \cref{lem:grid-corr-Lambdas} and the second follows from the fact that $\lambda \bar x + \epsilon \geq s$ for all $x \in A$.\footnote{Suppose first that $\lambda>0$. Then $\bar x\ge c_0/2>0$ on $A$, so $\lambda\bar x+\epsilon\ge\epsilon$, which is at least $s$ if $\lambda_\star\le0$. And, if $\lambda_\star>0$ then $\lambda\ge\lambda_\star$ gives $\lambda\bar x+\epsilon\ge\lambda_\star c_0/2+\epsilon\ge s$. Suppose instead that $\lambda\le0$, which can occur only if $\lambda_\star\le0$. Since $\bar x\le\tfrac12$ for every $x\in\Omega^{n}$, multiplying by $\lambda\le0$ gives $\lambda\bar x\ge\lambda/2\ge\lambda_\star/2$, and hence $\lambda\bar x+\epsilon\ge\epsilon+\lambda_\star/2\ge s$.} Since $s'\le\tfrac14$, for every $y\in\Omega$ we have $\Lambda(y+s')\ge\Lambda(-\tfrac12+s')=s'\ge s/2>0$. Therefore
\[
\Lambda(x_i+s')-\tfrac{2}{n^{2}}=\Lambda(x_i+s')\left[1-\frac{2/n^{2}}{\Lambda(x_i+s')}\right] \geq \Lambda(x_i+s') \left( 1-\tfrac{4}{sn^{2}} \right) .
\]
Using $\alpha_i \leq M$, it follows that
\[
\prod_{i\in [n]}\Lambda_{n,\m}\bigl(x_i+\lambda\bar x+\epsilon\bigr)^{\alpha_i-1}
\geq \left(1-\tfrac{4}{sn^{2}}\right)^{(M-1)n}\prod_{i\in [n]}\Lambda(x_i+s')^{\alpha_i-1}
\geq \tfrac12\prod_{i\in [n]}\Lambda(x_i+s')^{\alpha_i-1}
\]
for all large $n$, since $\bigl(1-\tfrac{4}{sn^{2}}\bigr)^{(M-1)n}\to1$. Hence, for all large $n$,
\begin{align*}
\mu_{n,\m}&=\Bigl(\prod_{i\in[n]}  \alpha_i\Bigr) \cdot \int_{\Omega^{n}} \prod_{i\in[n]}\Lambda_{n,\m}\bigl(x_i+\lambda \bar x +\epsilon\bigr)^{\alpha_i-1}\,dx \\
&\geq \Bigl(\prod_{i\in[n]}  \alpha_i\Bigr) \cdot \int_{A} \prod_{i\in[n]}\Lambda_{n,\m}\bigl(x_i+\lambda \bar x +\epsilon\bigr)^{\alpha_i-1}\,dx\\
&\geq \tfrac{1}{2} \cdot\Bigl(\prod_{i\in[n]}  \alpha_i\Bigr) \cdot \int_{A} \prod_{i\in [n]}\Lambda(x_i+s')^{\alpha_i-1}\,dx \\
&= \tfrac{1}{2} \cdot\Bigl(\prod_{i\in[n]}  \alpha_i\Bigr)\int_{\Omega^n} \prod_{i\in [n]}\Lambda(x_i+s')^{\alpha_i-1}\,dx \cdot \frac{\int_{A} \prod_{i\in [n]}\Lambda(x_i+s')^{\alpha_i-1}\,dx}{\int_{\Omega^n} \prod_{i\in [n]}\Lambda(x_i+s')^{\alpha_i-1}\,dx} \\
&= \tfrac{1}{2}\cdot\prod_{i\in[n]}\Bigl(\alpha_i\int_{\Omega}\Lambda(y+s')^{\alpha_i-1}\,dy\Bigr) \cdot \frac{\int_{A}\prod_{i\in[n]}\Lambda(x_i+s')^{\alpha_i-1}\,dx} {\int_{\Omega^{n}}\prod_{i\in[n]}\Lambda(x_i+s')^{\alpha_i-1}\,dx}\\
&\geq \tfrac{1}{2}\cdot(1+s')^{n}\cdot\tfrac12 \ \geq\ \tfrac14\bigl(1+\tfrac s2\bigr)^{n}\ \geq\ \bigl(1+\tfrac s4\bigr)^{n} .
\end{align*}

The first inequality restricts to $A$ and the second inequality follows from the argument above. The last line follows from \cref{lem:grid-corr-Lambdas} \ref{it:mass} and \ref{it:prob}, and the fact that $\tfrac{1}{4^{1/n}}(1+\tfrac s2) \geq 1+\tfrac s4$ for $n$ sufficiently large. This completes the proof of (i).

\bigskip
\noindent Case (ii): $\eta_n(\lambda)\geq s/2$. 
If $\lambda\le0$ then $|\lambda|\le1$, so $\eta_n(\lambda)\le\sqrt{\ln n/n}<s/2$ for all large $n$ and this case cannot arise. We may therefore assume that $\lambda>0$, so that, by the hypothesis of this case, $\lambda\ge\tfrac s2\sqrt{n/\ln n}$. For any $\mathbf a$ we write $\bar v (\mathbf a):= \tfrac 1n \sum_{j \in [n]} v_j(\mathbf a)$. Now fix $\mathbf a\in\mathbf A$. Since $v_i(\mathbf a)$ and $v_i(\mathbf b)$ both lie in the interval $\Omega$ of length $1$, we have $v_i(\mathbf a)-v_i(\mathbf b)\ge-1$ for every $i$ and every $\mathbf b$. Therefore, on the event
\[
E_{\mathbf a}:=\Bigl\{\ \bar v(\mathbf a)\ \ge\ \bar v(\mathbf b)+\tfrac1\lambda\text{ for every }\mathbf b\in B_1(\mathbf a)\setminus\{\mathbf a\}\ \Bigr\},
\]
we have, for every $i\in[n]$ and every such $\mathbf b$,
\[
u_i(\mathbf a)-u_i(\mathbf b)=\bigl[v_i(\mathbf a)-v_i(\mathbf b)\bigr]+\lambda\bigl[\bar v(\mathbf a)-\bar v(\mathbf b)\bigr]\ \ge\ -1+\lambda\cdot\tfrac1\lambda\ =\ 0 .
\]
Thus on $E_{\mathbf a}$ no agent can gain by deviating at all, so $\mathbf a$ is a pure Nash equilibrium and a fortiori a pure $\epsilon$-equilibrium for every $\epsilon\ge0$.
Hence
\[
p_{n,\m} := \mathbb E[I_{\mathbf a}]\ge\Pr[E_{\mathbf a}] \geq \Pr[\bar v(\mathbf a) \geq t + \tfrac1\lambda] \cdot \prod_{\mathbf b \in B_1(\mathbf a) \setminus \{\mathbf a\}} \Pr[\bar v(\mathbf b) \leq t],
\]
where the last inequality holds for any $t\geq 0$ and uses the fact that value averages $\bar v$ are independent across action profiles. 

Letting $K:= |B_1(\mathbf a) \setminus \{\mathbf a\}| \in \{n,\dots,(M-1)n\}$ and $t:= \sqrt{\ln K/(2n)}$, we now lower bound each of the terms of $p_{n,\m}$. Hoeffding's inequality gives $\Pr[\bar v(\mathbf b) \geq t] \leq e^{-2nt^2}$, hence $\Pr[\bar v(\mathbf b) \leq t] \geq 1- \Pr[\bar v(\mathbf b) \leq t]\geq 1 -e^{-2nt^2}$, and therefore $\prod_{\mathbf b \in B_1(\mathbf a) \setminus \{\mathbf a\}} \Pr[\bar v(\mathbf b) \leq t] \geq (1 - e^{-2nt^2})^K = (1-\tfrac1K)^K \geq \tfrac14$. Turning to the remaining term, let $u := t + \tfrac1\lambda$. Since $\lambda\ge\tfrac s2\sqrt{n/\ln n}$ and $K\le (M-1)n$, we have $u\le C\sqrt{\ln n/n}$ with $C:=\sqrt{(M-1)/2}+2/s$, so in particular $u<\tfrac14$ for all large $n$. Hence \cref{lem:corr-tails}, together with the fact that $1-x\ge e^{-2x}$ for $x\in[0,\tfrac12]$, gives $\Pr\bigl[\bar v(\mathbf a)\ge u\bigr] \geq \tfrac12(1-2u)^{n} \geq \tfrac12e^{-4un} \geq \tfrac12e^{-4C\sqrt{n\ln n}}$. Combining our bounds, it follows that for all large $n$, uniformly in $\m$,
\[
\mu_{n,\m} \geq 2^n p_{n,\m} \geq \tfrac{2^n}{8} e^{-4C \sqrt{n \ln n}}  \geq (1 + \tfrac s4)^n,
\]
where the last inequality holds for all large $n$ because $s \leq \tfrac 14$.
\end{proof}

\begin{proof}[Proof of \cref{lem:grid-corr-p-bound}]
Write $\eta_n(\lambda):=|\lambda|\sqrt{\ln n/n}$, $\lambda_1:=50M$, $\bar x:=\tfrac1n\sum_{i \in [n]} x_i$, and set $\beta:=\tfrac{1-\epsilon}{2\lambda_1}>0$ and $A:=\{x\in\Omega^{n}:\lambda\bar x\le\lambda_1\beta\}$. Since $\mathcal L(n) \subseteq [-1,\infty)$ and $\lambda \leq \lambda_1$, we have $|\lambda| \leq \lambda_1$ throughout. By Part (a) of \cref{prop:grid-corr}, $p_{n,\m}=\int_{\Omega^n}\prod_{i \in [n]}\Lambda_{n,\m}(x_i+\lambda\bar x+\epsilon)^{\alpha_i-1}dx$, and we bound the contributions of $A$ and $A^{c}$ separately.

\medskip
\noindent Contribution of $A$. Since $\alpha_i\ge2$ and $\lambda\bar x\le\lambda_1\beta=\tfrac{1-\epsilon}{2}$, letting $s_1:=\tfrac{1+\epsilon}{2}$, we have that for every $x\in A$,
\[
\prod_{i \in [n]}\Lambda_{n,\m}\bigl(x_i+\lambda\bar x+\epsilon\bigr)^{\alpha_i-1}  \leq \prod_{i\in [n]}\Lambda_{n,\m}\bigl(x_i+s_1\bigr) \leq \prod_{i\in [n]}\Lambda\bigl(x_i+s_1+\eta_n(\lambda)\bigr),
\]
where the last step follows from \cref{lem:grid-corr-Lambdas}\ref{it:sandwich}.  Suppose $n$ is large enough that $\eta_n(\lambda_1)\le\tfrac{1-\epsilon}{8}$, so that
$s_1+\eta_n(\lambda)\le s_2:=\tfrac{5+3\epsilon}{8}<1$. Enlarging the domain from $A$ to
$\Omega^{n}$ and using that the bound factorizes across coordinates,
\[
\int_A\prod_{i\in[n]}\Lambda_{n,\m}\bigl(x_i+\lambda\bar x+\epsilon\bigr)^{\alpha_i-1}dx
\ \le\ \Bigl(\int_\Omega\Bigl[\Lambda(x+s_2)+\tfrac{2}{n^{2}}\Bigr]dx\Bigr)^{n}
\ =\ \Bigl(1-\tfrac{9(1-\epsilon)^{2}}{128}+\tfrac{2}{n^{2}}\Bigr)^{n}
\ \le\ \Bigl(1-\tfrac{(1-\epsilon)^{2}}{32}\Bigr)^{n} ,
\]
where the equality follows from \cref{lem:grid-corr-Lambdas}\ref{it:mass} and the last step holds for all large $n$ because $\tfrac{9}{128}>\tfrac{1}{32}$.

\medskip
\noindent Contribution of $A^c$. Here $\lambda\bar x>\lambda_1\beta\ge|\lambda|\beta$, which
forces $\lambda\ne0$ and $|\bar x|>\beta$. Bounding the integrand by $1$ and
applying Hoeffding's inequality to $\bar X:= \tfrac1n \sum_{i \in [n]} X_i$ where the $X_i$ are iid variables that are uniform in $\Omega$,
\[
\int_{A^{c}}\prod_{i \in [n]}\Lambda_{n,\m}\bigl(x_i+\lambda\bar x+\epsilon\bigr)^{\alpha_i-1}dx
\ \le\ \bigl|A^{c}\bigr|\ \le\ \Pr\bigl[|\bar X|>\beta\bigr]\ \le\ 2e^{-2n\beta^{2}}
=2\bigl(e^{-(1-\epsilon)^{2}/(2\lambda_1^{2})}\bigr)^{n}.
\]

Adding the two contributions gives $p_{n,\m}\le\bar q^{\,n}+2\bar q^{\,n}=3\bar q^{\,n}$.
\end{proof}

\begin{proof}[Proof of \cref{lem:var-lower-bound}]
The proof of this lemma leverages the geometry of $\mathbf A$, careful conditioning, and the fact that the dependence range in $(I_{\mathbf a})_{\mathbf a\in\mathbf A}$ is $r_{n,\m}=2$.

For any $\mathbf y \in \mathbf Y :=\prod_{i \in [n]\setminus\{1\}}[\alpha_i]$, we call the set of profiles $L(\mathbf y) := \bigl\{ (x,\mathbf y) : x \in [\alpha_1] \bigr\} $ a \emph{line}, and $\mathcal F := \{L(\mathbf y) : \mathbf y \in \mathbf Y\}$ is the collection of all lines. All lines are taken in direction $1$; nothing below depends on that choice. The \emph{one-step neighborhood} of $L \in \mathcal F$ is $\mathcal N(L) := \bigcup_{\mathbf a \in L} B_1(\mathbf a)$; these are the profiles of $L$ together with those at Hamming distance $1$ from a profile in $L$. Observe that $|\mathcal N(L)| \leq nM^{2}$, and that every $\mathbf a \in \mathcal N(L)\setminus L$ is at distance $1$ from exactly one profile in $L$. Call $\mathcal D \subseteq \mathcal F$ \emph{separated} if $d_H(\mathbf a,\mathbf a')\ge 3$ for all distinct $L,L' \in \mathcal D$ and all $\mathbf a \in L$, $\mathbf a' \in L'$. Observe that a separated $\mathcal D$ with $|\mathcal D| \ge |\mathbf A|/(nM)^{3} $ always exists.\footnote{We must select points in $\mathbf Y$ so that $d_H(\mathbf y,\mathbf y')\ge3$. Choose them greedily, accepting any point at distance at least $3$ from all those already accepted. When none remains, the balls of radius $2$ about the accepted points cover $\mathbf Y$, so at least $|\mathbf Y|/\beta_{\mathbf Y}(2)$ were accepted. Since $|\mathbf Y| = |\mathbf A|/\alpha_1 \ge |\mathbf A|/M$ and $\beta_{\mathbf Y}(2) \le \beta_{\mathbf A}(1)^{2} \le (nM)^{2}$ by \cref{lem:ball-power}, the claim follows.} 

Fix a separated $\mathcal D$. For any $L \in \mathcal D$, let $W_L := \sum_{\mathbf a \in \mathcal N(L)} I_{\mathbf a}$ denote the number of pure $\epsilon$-equilibria among the profiles in $\mathcal N(L)$, and let $R:=\sum_{\mathbf a\notin\bigcup_{L}\mathcal N(L)}I_{\mathbf a}$ denote the number of pure $\epsilon$-equilibria at a Hamming distance of at least 2 from any line in $\mathcal D$. Then we have the following identity:
\[
S_{n,\m}= \sum_{\mathbf a\in\mathbf A} I_{\mathbf a} = R + \sum_{L\in\mathcal D} W_L .
\]
Let $Y:= (v_i(\mathbf a))_{i \in [n], \mathbf a \not\in \bigcup_{L\in\mathcal D}L}$; this is the list of all values at profiles that are not in any line, and let $F_Y$ denote the distribution of $Y$. For any $L \in \mathcal D$, let $Z_L:=(v_i(\mathbf a))_{i \in [n], \mathbf a \in L}$; this is the list of all values at profiles in line $L$. The variables $R$ and $W_L$ depend on $Y$ and $Z_L$ in the following manner: (a) if $\mathbf a \notin \bigcup_{L\in\mathcal D}\mathcal N(L)$ then $B_1(\mathbf a)\cap L = \emptyset$ for every $L \in \mathcal D$, so $R$ depends on $Y$ alone; and (b) if $\mathbf a \in \mathcal N(L)$ then $B_1(\mathbf a)\cap L' = \emptyset$ for every $L' \in \mathcal D$ with $L'\ne L$. Hence $W_L$ depends on $Y$ and $Z_L$ only. This allows us to write:
\[
  \operatorname{Var}[S_{n,\m}]
  = \mathbb E\bigl[\operatorname{Var}[S_{n,\m}\mid Y]\bigr]
     + \operatorname{Var}\bigl[\mathbb E[S_{n,\m}\mid Y]\bigr] 
  \ge \mathbb E\bigl[\operatorname{Var}[S_{n,\m}\mid Y]\bigr] 
  = \sum_{L\in\mathcal D}\mathbb E\bigl[\operatorname{Var}[W_L\mid Y]\bigr] .
\]
The first equality is the law of total variance, and the inequality follows because the discarded term is non-negative. For the final equality, note that $Y$ and the lists $Z_L$, $L\in\mathcal D$, are mutually independent; consequently, conditional on $Y$, the $Z_L$ remain mutually independent. Since (a) $R$ is a function of $Y$ alone, it is constant conditional on $Y$; and (b) since each $W_L$ is a function of $Y$ and $Z_L$ only, the variables $W_L$, $L\in\mathcal D$, are conditionally independent given $Y$, which gives $ \operatorname{Var}[\sum_{L\in\mathcal D}W_L \mid Y] =\sum_{L\in\mathcal D}\operatorname{Var}[W_L\mid Y]$.

We now turn to lower bounding $\mathbb E\bigl[\operatorname{Var}[W_L\mid Y]\bigr]$. To do so, for each $L \in \mathcal D$, let $Y_L := \bigl(v_i(\mathbf a)\bigr)_{i\in[n],\ \mathbf a\in\mathcal N(L)\setminus L}$ denote the list of values at the profiles in the one-step neighborhood of $L$ other than those of $L$ itself (so $Y_L$ is a sub-list of $Y$), and define the set
\[
\mathcal T_L:=\left\{(y_i(\mathbf a))_{i \in [n], \mathbf a \in (\mathcal N(L) \setminus L)} \in \Omega^{[n]\times (\mathcal N(L)\setminus L)}: \,\Big|\tfrac1n\sum_{j \in [n]}y_j(\mathbf a)\Big|\le\varpi_n \ \text{ for every } \mathbf a \in \mathcal N(L)\setminus L \,\right\}
\]
where $\varpi_n:=\sqrt{\ln(4nM^{2})/(2n)}$. Each $\bar v(\mathbf a):=\tfrac1n\sum_{j\in[n]}v_j(\mathbf a)$ is an average of $n$ independent values of mean $0$ and range $1$, so Hoeffding's inequality gives $\Pr\bigl[|\bar v(\mathbf a)|>\varpi_n\bigr]\le 2e^{-2n\varpi_n^{2}}=\tfrac{1}{2nM^{2}}$ and, since $|\mathcal N(L)\setminus L|\le|\mathcal N(L)|\le nM^{2}$, a union bound gives $\Pr[Y_L \in \mathcal T_L]\ge\tfrac12$. 

Fix $\delta := \tfrac{1-\epsilon}{4}$ and, for each $L \in \mathcal D$, define the set
\[
\mathcal S_L:= \left\{ (z_i(\mathbf a))_{i \in [n], \mathbf a \in L} \in \Omega^{[n] \times L} : \exists \mathbf a^* \in \arg\max_{\mathbf a \in L} \sum_{j=2}^n z_j(\mathbf a) \text{ such that}
\left\{\begin{matrix}
z_1(\mathbf a^*) \geq \tfrac12 - \delta \text{ and } \\
\max_{\mathbf a \in L \setminus \{\mathbf a^*\}} z_1(\mathbf a) \leq \delta -\tfrac12 
\end{matrix}\right\}
\right\}
\]
The key property of this set is that if $z \in \mathcal S_L$ and $\mathbf a^*$ is the maximizer (for set membership) then for every $\mathbf a \in L \setminus \{\mathbf a^*\}$, 
\[
u_1(\mathbf a^*) - u_1(\mathbf a) = (1+ \tfrac \lambda n)(z_1(\mathbf a^*) - z_1(\mathbf a) )+ \tfrac \lambda n \sum_{j =2}^n (z_j(\mathbf a^*) - z_j(\mathbf a)) \geq 1-2\delta = \frac{1+\epsilon}{2} > \epsilon.
\]
Thus at most one profile of $L$ is a pure $\epsilon$-equilibrium, namely $\mathbf a^{*}$; and the witness $\mathbf a^{*}$ is itself unique since $\tfrac12-\delta>\delta-\tfrac12$ forbids two profiles from simultaneously meeting the first condition and the second. Now, note that a maximizer $\mathbf a^*$ of $\sum_{j=2}^{n}v_j(\mathbf a)$ is determined by the coordinates $(v_j(\mathbf a))_{j\ge2, \mathbf a\in L}$, which are independent of the coordinates $(v_1(\mathbf a))_{\mathbf a\in L}$. Hence,
\[
\Pr\bigl[Z_L\in\mathcal S_L\bigr]\ \ge\ \sum_{\mathbf b\in L}\Pr[\mathbf a^*=\mathbf b]\cdot \Pr\Bigl[v_1(\mathbf b)\ge\tfrac12-\delta\ \text{ and} \max_{\mathbf a\in L\setminus\{\mathbf b\}}v_1(\mathbf a)\le\delta-\tfrac12\Bigr]
\ =\ \delta^{\alpha_1}.
\]
Finally, letting $t_*:= \tfrac 32 + \epsilon + \lambda \varpi_n$, and each $L \in \mathcal D$ and $\mathbf b \in L$, define
\[
\mathcal M_L^{\pm}(\mathbf b):= \left\{(z_i(\mathbf a))_{i \in [n], \mathbf a \in L} \in \Omega^{[n] \times L} : \pm \tfrac{\lambda}{n} \sum_{j=2}^n z_j(\mathbf b) \geq t_* + \tfrac{\lambda}{2n} \right\},
\]
and for each $H \subseteq L$,
\[
\mathcal E_L^H := \mathcal S_L \cap \bigcap_{\mathbf a \in H} \mathcal M_L^+(\mathbf a) \cap \bigcap_{\mathbf a \in L \setminus H} \mathcal M_L^-(\mathbf a) .
\]
So $\mathcal E_L^H $ is a set of value lists that result in at most one pure $\epsilon$-equilibrium in $L$ (via $\mathcal S_L$), high aggregate values for agents other than 1 at profiles in $H$ (via $\mathcal M_L^+$), and low aggregate values for agents other than 1 at profiles in $L \setminus H$ (via $\mathcal M_L^-$). By symmetry of the uniform distribution, $\Pr[Z_L \in  \mathcal M_L^+(\mathbf a)] = \Pr[Z_L \in \mathcal M_L^-(\mathbf a)]=:\nu_{n,\m}$, and, by the same conditioning argument as above, we obtain $\Pr[Z_L \in \mathcal E_L^H] \geq (\delta \nu_{n,\m})^{\alpha_1} \geq (\delta \nu_{n,\m})^{M}$.

Recall that $W_L$ depends on $Z_L$ and $Y$ only. We may therefore write $W_L=w(Z_L,Y)$ and 
\[
\mathbb E\bigl[\operatorname{Var}[W_L\mid Y]\bigr]  \geq \mathbb E\Bigl[\mathbbm 1\{Y_L\in\mathcal T_L\}\operatorname{Var}\bigl[W_L\mid Y\bigr]\Bigr] = \int_{\{y\,:\,y_L\in\mathcal T_L\}}\operatorname{Var}\bigl[w(Z_L,y)\bigr]\,dF_Y(y).
\]
We now use the facts above to lower bound $\operatorname{Var}\bigl[w(Z_L,y)\bigr]$. To this end, let $y$ be a realization of $Y$ with $y_L\in\mathcal T_L$. We claim that
\begin{enumerate}[(i),topsep=0ex,itemsep=0ex]
    \item $w(z,y)=1$ for any $z\in\mathcal E^{L}_L$,
    \item there is a constant $c_y$ such that $w(z,y)=c_y$ for any  $z\in\mathcal E^{\emptyset}_L$,
    \item if $c_y\geq 1$ then there is $\mathbf a\in L$ with $w(z,y)\geq 2$ for any $z\in\mathcal E^{\{\mathbf a\}}_L$.
\end{enumerate}
One can verify that if $y_L \in \mathcal T_L$ then for every $H \subseteq L$, $z \in \mathcal E_L^H$, $i \in [n]$, $\mathbf a \in L$, and $\mathbf b \in \mathcal N(L) \setminus L$,\footnote{Indeed, let $\mathbf a \in H$ and $z \in \mathcal M_L^+(\mathbf a)$. Since $z_1(\mathbf a)\geq -\tfrac12$, we have $\tfrac{\lambda}{n}\sum_{j=2}^{n}z_j(\mathbf a)+\tfrac{\lambda}{n}z_1(\mathbf a) \geq t_*+\tfrac{\lambda}{2n}-\tfrac{\lambda}{2n}=t_*$, and therefore $u_i(\mathbf a)\geq t_* - \tfrac12$ for each $i \in [n]$. Symmetrically, if $\mathbf a\in L\setminus H$ then $u_i(\mathbf a)\le\tfrac12-t_*$ for every $i$. Moreover, for any $\mathbf b \in \mathcal N(L)\setminus L$, since $y_L \in \mathcal T_L$, we have $|u_i(\mathbf b)| \leq \tfrac12 + \lambda \varpi_n$ for every $i$. Putting these together gives us $(\dagger)$ and $(\ddagger)$.}
\begin{enumerate}[topsep=0ex,itemsep=0ex]
    \item[$(\dagger)$] $\mathbf a \in H$ implies $u_i(\mathbf a) - u_i(\mathbf b) \geq \tfrac12 + \epsilon >\epsilon$,
    \item[$(\ddagger)$] $\mathbf a \not\in H$ implies $u_i(\mathbf b) - u_i(\mathbf a) \geq \tfrac12 + \epsilon >\epsilon$.
\end{enumerate}

To prove (i), suppose $H=L$. Then every $\mathbf b \in \mathcal N(L) \setminus L$ has a neighbor $\mathbf a \in H$ so no such $\mathbf b$ can be a pure $\epsilon$-equilibrium by $(\dagger)$; however, some profile in $L$ must be a pure $\epsilon$-equilibrium: since $z \in \mathcal S_L$, agent 1 meets the conditions at some $\mathbf a^*$, and any neighbor $\mathbf b \in \mathcal N(L)\setminus L$ of $\mathbf a^*$ must satisfy $u_j(\mathbf a^*) - u_j(\mathbf b) > \epsilon$ by $(\dagger)$. Since there can be at most one pure $\epsilon$-equilibrium in $L$, $w(z,y)=1$ for every $z\in\mathcal E^{L}_L$.

To prove (ii), suppose $H=\emptyset$. Then $(\ddagger)$ implies that no profile in $L$ can be a pure $\epsilon$-equilibrium. Now let $\mathbf b\in\mathcal N(L)\setminus L$, let $\mathbf a\in L$ be its unique neighbor in $L$, and let $i\neq1$ be the coordinate in which $\mathbf a$ and $\mathbf b$ differ. Whether $\mathbf b$ is a pure $\epsilon$-equilibrium is determined by the values at the profiles in $B_1(\mathbf b)$, and $\mathbf a$ is the only one of these that lies on a line of $\mathcal D$. The values at $\mathbf a$ enter only agent $i$'s condition at $\mathbf b$, through the comparison $u_i(\mathbf b)\ge u_i(\mathbf a)-\epsilon$, which holds automatically since $(\ddagger)$ gives $u_i(\mathbf b)-u_i(\mathbf a)>\epsilon$. Hence, on $\mathcal E^{\emptyset}_L$, whether $\mathbf b$ is a pure $\epsilon$-equilibrium is determined by $y$, and so there is some $c_y$ such that $w(z,y)=c_y$ for every $z\in\mathcal E^{\emptyset}_L$.

To prove (iii), suppose $c_y\ge1$. By the argument for (ii), $c_y$ counts pure $\epsilon$-equilibria of $\mathcal N(L)\setminus L$, so there is $\mathbf b_0\in\mathcal N(L)\setminus L$ for which $I_{\mathbf b_0}=1$ on $\mathcal E^{\emptyset}_L$; let $\mathbf c_0$ be its unique neighbor in $L$ and choose $\mathbf a\in L\setminus\{\mathbf c_0\}$. Let $z\in\mathcal E^{\{\mathbf a\}}_L$. Since $\lambda>0$, the conditions $\mathcal M^{+}_L(\mathbf a)$ and $\mathcal M^{-}_L(\mathbf a')$ for $\mathbf a'\in L\setminus\{\mathbf a\}$ force $\sum_{j=2}^{n}z_j(\mathbf a)>0>\sum_{j=2}^{n}z_j(\mathbf a')$, so the witness for membership in $\mathcal S_L$ is $\mathbf a^{*}=\mathbf a$. Given this, $(\dagger)$ and $z\in\mathcal S_L$ imply that $\mathbf a$ is the unique pure $\epsilon$-equilibrium in $L$. Since $\mathbf c_0\notin\{\mathbf a\}=H$, $(\ddagger)$ still gives $u_i(\mathbf b_0)-u_i(\mathbf c_0)>\epsilon$ for the coordinate $i$ in which $\mathbf b_0$ and $\mathbf c_0$ differ, so, by the argument of (ii), whether $\mathbf b_0$ is a pure $\epsilon$-equilibrium is determined by $y$ exactly as on $\mathcal E^{\emptyset}_L$. Hence $I_{\mathbf b_0}=1$. Since the profiles $\mathbf a$ and $\mathbf b_0$ are distinct, $w(z,y)\ge2$.

Fix $y$ with $y_L\in\mathcal T_L$. Since $Y$ and $Z_L$ are independent, $\Pr\bigl[Z_L\in\mathcal E^{H}_L\mid Y=y\bigr]=\Pr\bigl[Z_L\in\mathcal E^{H}_L\bigr] \ge(\delta\nu_{n,\m})^{M}$ for every $H\subseteq L$. If $c_y=0$, apply \cref{lem:sep} with $k=0$, $A:=\mathcal E^{\emptyset}_L$ and $B:=\mathcal E^{L}_L$: by (ii) we have $w(z,y)=0\le k$ for every $z\in A$, and by (i) we have $w(z,y)=1\ge k+1$ for every $z\in B$. If instead $c_y\ge1$, apply \cref{lem:sep} with $k=1$, $A:=\mathcal E^{L}_L$ and $B:=\mathcal E^{\{\mathbf a\}}_L$ for the profile $\mathbf a\in L$ supplied by (iii): by (i) we have $w(z,y)=1\le k$ for every $z\in A$, and by (iii) we have $w(z,y)\ge2\ge k+1$ for every $z\in B$. In either case \cref{lem:sep} gives $\operatorname{Var}\bigl[w(Z_L,y)\bigr]\ge\tfrac14(\delta\nu_{n,\m})^{M}$. We therefore have that
\[
\mathbb E\bigl[\operatorname{Var}[W_L\mid Y]\bigr] \geq \int_{\{y\,:\,y_L\in\mathcal T_L\}}\operatorname{Var}\bigl[w(Z_L,y)\bigr]\,dF_Y(y) \geq \Pr[Y_L \in \mathcal T_L] \tfrac{1}{4}(\delta \nu_{n,\m})^M \geq \tfrac{1}{8}(\delta \nu_{n,\m})^M ,
\]
and therefore $\operatorname{Var}[S_{n,\m}] \geq \sum_{L \in \mathcal D} \mathbb E\bigl[\operatorname{Var}[W_L\mid Y]\bigr] \geq |\mathcal D| \tfrac{1}{8}(\delta \nu_{n,\m})^M \geq \frac{|\mathbf A|}{(nM)^3}  \tfrac18 (\delta \nu_{n,\m})^M$.

To complete the proof, it remains for us to establish a lower bound on $\tfrac18 (\delta \nu_{n,\m})^M$. Let $\bar v_{-1}(\mathbf a):=\tfrac{1}{n-1}\sum_{i=2}^{n}v_i(\mathbf a)$, so that $\tfrac{\lambda}{n}\sum_{j=2}^{n}v_j(\mathbf a)=\tfrac{\lambda(n-1)}{n}\bar v_{-1}(\mathbf a)$ and the condition defining $\mathcal M^{+}_L(\mathbf a)$ reads $\bar v_{-1}(\mathbf a)\ge \varrho_n$, where $\varrho_n:=\frac{n}{n-1}(\frac{t_*}{\lambda}+\frac{1}{2n}) =\frac{n}{n-1}(\frac{\epsilon + 3/2}{\lambda}+\varpi_n+\frac{1}{2n}) \leq \frac54 (\frac{5}{2\lambda}+\varpi_n+\frac{1}{2n} )$ for $n\ge5$, using $\epsilon<1$. Since $\lambda> 50M\ge100$ and $\varpi_n\to0$, in particular $\varrho_n<\tfrac14$ for all large $n$, uniformly in $\m$. We can therefore apply \cref{lem:corr-tails} and the fact that $1-x\ge e^{-2x}$ for $x\in[0,\tfrac12]$ to get
\[
\nu_{n,\m}=\Pr\bigl[Z_L\in\mathcal M^{+}_L(\mathbf a)\bigr]= \Pr\bigl[\bar v_{-1}(\mathbf a)\ge \varrho_n\bigr]  \geq \tfrac12(1-2 \varrho_n)^{n-1} \geq \tfrac12 e^{-4 \varrho_n(n-1)} = e^{-4 \varrho_n(n-1) - \ln 2}.
\]
Hence $\tfrac18 (\delta \nu_{n,\m})^M \geq e^{- \varkappa_{n,\m} n} $ where $\varkappa_{n,\m}:= \tfrac1n M(\ln 2 - \ln \delta) + \tfrac1n\ln 8 + \tfrac{n-1}{n}4M \varrho_n $. But since $\epsilon \leq 1$, $\tfrac{n-1}{n}4M \varrho_n \leq \tfrac{10 M}{\lambda} + 4M \varpi_n + \tfrac{2M}{n}$. So for $\lambda > 50M$, the first term is at most $1/5 < \tfrac{3}{10} \ln 2$ and the remaining terms vanish with $n$ large. Consequently, for all large $n$, uniformly in $\m$, $\tfrac18 (\delta \nu_{n,\m})^M \geq e^{- \varkappa_{n,\m} n} \geq e^{- n \tfrac{3}{10} \ln 2}$.
\end{proof}

\subsection{\texorpdfstring{Proof of \cref{thm:correlation-nonexistence}}{Proof for correlation across agents: non-existence}}\label{sec:thm-corr-nonexistence}

The utility model here is that of \cref{sec:correlation} with $\lambda=-1$. All that is needed is an \emph{upper} bound on $\mu_{n,\m}$, which the lemma below supplies. Its proof rests on two features of the zero-sum case: the utilities at any action profile sum to zero, and each of them is log-concavely distributed.

\begin{lemma}\label{lem:grid-corr-mu-upper}
Fix $\epsilon\ge0$, let $\mathbf A\in\mathcal A_M(n)$, let $\lambda=-1$, and write $\m=(\mathbf A,\lambda)$, $G_{\mathbf A}:=\bigl(\prod_{i\in[n]}\alpha_i\bigr)^{1/n}$ and $m_{\mathbf A}:=\min_{i\in[n]}\alpha_i$. Then
\[
\mu_{n,\m}\ \le\ \Bigl[G_{\mathbf A}\,\Lambda_{n,\m}(\epsilon)^{\,m_{\mathbf A}-1}\Bigr]^{n}
\ \le\ \Bigl[G_{\mathbf A}\bigl(\tfrac12+\tfrac{n}{n-1}\epsilon\bigr)^{\,m_{\mathbf A}-1}\Bigr]^{n}.
\]
\end{lemma}

\begin{proof}[Proof of \cref{lem:grid-corr-mu-upper}]
For any $\mathbf a\in\mathbf A$, the argument in the proof of \cref{prop:grid-corr} gives
\[
p_{n,\m}=\mathbb E\Bigl[\textstyle\prod_{i\in[n]}
\Lambda_{n,\m}\bigl(u_i(\mathbf a)+\epsilon\bigr)^{\alpha_i-1}\Bigr].
\]
Since $\Lambda_{n,\m}\le1$ and $\alpha_i-1\ge m_{\mathbf A}-1$ for every $i$, the term in the expectation is at most $\bigl(\prod_{i\in[n]}\Lambda_{n,\m}(u_i(\mathbf a)+\epsilon)\bigr)^{m_{\mathbf A}-1}$.

Since $\lambda=-1$ we have $u_i(\mathbf a)=\tfrac{n-1}{n}v_i(\mathbf a)-\tfrac1n\sum_{j\neq i}v_j(\mathbf a)$, a linear combination of independent uniform random variables. Uniform densities are log-concave, and log-concavity is preserved by scaling and convolution, so the density of $u_i(\mathbf a)$ is log-concave; hence so is its cdf $\Lambda_{n,\m}$.\footnote{See, e.g., \citet{bagnoli2005log}: if a density is log-concave then so is the associated distribution function.} Applying Jensen's inequality to the concave function $\ln\Lambda_{n,\m}$ therefore gives, pointwise,
\[
\tfrac1n\sum_{i\in[n]}\ln\Lambda_{n,\m}\bigl(u_i(\mathbf a)+\epsilon\bigr)
\ \le\ \ln\Lambda_{n,\m}\Bigl(\tfrac1n\textstyle\sum_{i\in[n]}u_i(\mathbf a)+\epsilon\Bigr)
\ =\ \ln\Lambda_{n,\m}(\epsilon),
\]
where the last step follows from $\sum_{i\in[n]}u_i(\mathbf a)=0$. Combining the two displays, $p_{n,\m}\le\Lambda_{n,\m}(\epsilon)^{\,n(m_{\mathbf A}-1)}$, and multiplying by $|\mathbf A|=\prod_{i\in[n]}\alpha_i=G_{\mathbf A}^{\,n}$ gives the first bound.

For the second bound, write $u_i(\mathbf a)=X+Y$ with $X:=\tfrac{n-1}{n}v_i(\mathbf a)$ and $Y:=-\tfrac1n\sum_{j\neq i}v_j(\mathbf a)$, which are independent, and let $f_X$ and $f_Y$ denote their densities. Then
\begin{align*}
\Lambda_{n,\m}(\epsilon)-\Lambda_{n,\m}(0)
&=\Pr\bigl[u_i(\mathbf a)\in(0,\epsilon)\bigr]=\Pr\bigl[X+Y\in(0,\epsilon)\bigr]\\
&=\int \Pr\bigl[X\in(-y,\epsilon-y)\bigr] f_Y(y)\,dy
=\int\Bigl(\int_{-y}^{\epsilon-y}f_X(x)\,dx\Bigr)f_Y(y)\,dy .
\end{align*}
Since $f_X$ equals $\tfrac{n}{n-1}$ on an interval of length $\tfrac{n-1}{n}$ and vanishes elsewhere, the inner integral is at most $\tfrac{n}{n-1}\epsilon$. Moreover, since the values $v_j(\mathbf a)$, $j\in[n]$, are independent and symmetric about zero, $u_i(\mathbf a)$ is symmetric about zero and, being continuously distributed, satisfies $\Lambda_{n,\m}(0)=\tfrac12$. Hence $\Lambda_{n,\m}(\epsilon)\le\tfrac12+\tfrac{n}{n-1}\epsilon$.
\end{proof}

\begin{proof}[Proof of \cref{thm:correlation-nonexistence}]
Fix any $\varsigma'\in(\varsigma,1)$. For every $\m=(\mathbf A,-1)\in\mathcal I(n)$ we have $G_{\mathbf A}\bigl(\tfrac12+\epsilon\bigr)^{m_{\mathbf A}-1}\le\varsigma$ by definition of $\mathcal A_M(n;\epsilon,\varsigma)$, and, using $\tfrac12+\epsilon\ge\tfrac12$ and $m_{\mathbf A}\le M$,
\[
\Bigl(\tfrac12+\tfrac{n}{n-1}\epsilon\Bigr)^{m_{\mathbf A}-1}
\ \le\ \bigl(\tfrac12+\epsilon\bigr)^{m_{\mathbf A}-1}
\Bigl(1+\tfrac{2\epsilon}{n-1}\Bigr)^{M-1} .
\]
The final factor depends only on $n$, $M$ and $\epsilon$ and tends to $1$, so there is an integer $\bar n$ such that, by \cref{lem:grid-corr-mu-upper},
\[
\mu_{n,\m}\ \le\ \Bigl[G_{\mathbf A}\bigl(\tfrac12+\epsilon\bigr)^{m_{\mathbf A}-1}
\Bigl(1+\tfrac{2\epsilon}{n-1}\Bigr)^{M-1}\Bigr]^{n}
\ \le\ (\varsigma')^{\,n}
\]
for all $n\ge \bar n$ and all $\m\in\mathcal I(n)$. Since $S_{n,\m}$ is a non-negative integer-valued random variable, Markov's inequality gives
$\Pr[S_{n,\m}\ge1]\le\mathbb E[S_{n,\m}]=\mu_{n,\m}$. Hence $1-\mathbb P^{\epsilon}_{n,\m}(0)=\Pr[S_{n,\m}\ge1]\le(\varsigma')^{\,n}$ for all $n\ge \bar n$ and all $\m\in\mathcal I(n)$, and therefore $\mathbb P^{\epsilon}_{n,\m}(0)\Longrightarrow1$.
\end{proof}

\begin{remark*}
The conclusion also follows from the \cref{thm:master} Part \ref{thm:master:poisson}: \cref{prop:grid-corr} Part (a) gives $\theta_{n,\m}=(nM)^2$ and $b_{n,\m}\le(nM)^2$, and since $|\mathbf A|\ge2^{n}$ the bound above yields $\mu_{n,\m}\bigl(b_{n,\m}p_{n,\m}+\theta_{n,\m}\bigr)\le2(nM)^{2}(\varsigma')^{\,n} \Longrightarrow0$, so that $\mathbb P^{\epsilon}_{n,\m}\Longrightarrow\Pi_0$. This is more than is needed here: unlike our other results, the present one turns on the mean alone, and Markov's inequality suffices.
\end{remark*}

\subsection{\texorpdfstring{Proof of \cref{thm:graphical}}{Proof for correlation across action profiles}}

\begin{proposition}[Correlation across action profiles]\label{prop:grid-graph}
Fix $\epsilon\ge0$, $c>0$, an integer $M\ge2$, and a continuous distribution $F$ on $\mathbb R$. For each $n\ge2$ and each $\m = (\mathbf A,G, F)\in\mathcal I(n) = \mathcal A_M(n) \times \mathcal G_c(n) \times \{F\}$, let $I_{\mathbf a}$ be the indicator that $\mathbf a$ is a pure $\epsilon$-equilibrium of $g=(\mathbf A, \{u_i\}_{i \in [n]})$, where $g$ has utilities drawn from $\mathbb J_{n,\m}$ as described in \cref{sec:graphical}. Then we have:
\begin{enumerate}[(a),topsep=0ex,itemsep=0ex,leftmargin=0cm]
\item $(I_{\mathbf a})_{\mathbf a\in\mathbf A}$ is a $(r_{n,\m},\theta_{n,\m})$-well-behaved indicator grid for all sufficiently large $n$ with
\[
r_{n,\m}=\Bigl\lceil\frac{n}{c\ln n}\Bigr\rceil-1 \text{ and }  \theta_{n,\m}:=(e^{x_n}-1) \text{ where } x_n:=M \,n^{\,1-c\ln(1/\xi_{F,2}^\epsilon)}.
\]
Moreover, letting $S_{n,\m}:=\sum_{\mathbf a\in\mathbf A}I_{\mathbf a}$, $\mu_{n,\m}:=\mathbb E[S_{n,\m}]$, and $p_{n,\m}:=\frac{\mu_{n,\m}}{|\mathbf A|}$, we have
\[
\mu_{n,\m}=\prod_{i\in[n]}\alpha_i \,\xi_{F,\alpha_i}^\epsilon, \text{ and } b_{n,\m}:= \beta_{\mathbf A}(r_{n,\m}) \leq \left( \frac{e n M}{r_{n,\m}}\right)^{r_{n,\m}}.
\]
\item If $\epsilon =0 $ and $c > 1/\ln 2$ then the hypotheses of the \cref{thm:master} Part \ref{thm:master:poisson} are satisfied; and if $\epsilon > 0 $ and $c > 1/\ln (1/\xi_{F,2}^\epsilon)$ then the hypotheses of the \cref{thm:master} Parts \ref{thm:master:existence} and \ref{thm:master:clt} are satisfied.
\end{enumerate}
\end{proposition}
\noindent\cref{thm:graphical} follows directly from \cref{prop:grid-graph} and the \cref{thm:master}.\hfill$\Box$

Observe that, in this case, the dependence range $r_{n,\m}$ grows with $n$.

\begin{proof}[Proof of \cref{prop:grid-graph}]
    For $\mathbf a,\mathbf a'\in\mathbf A \in \mathcal A_M(n)$, let $\Delta(\mathbf a,\mathbf a'):=\{i\in[n]:a_i\neq a'_i\}$ denote the set of coordinates at which the two profiles differ, so that $d_H(\mathbf a,\mathbf a')=|\Delta(\mathbf a,\mathbf a')|$. Suppose $G \in \mathcal G_c(n)$ is a $\gamma$-expander with $\gamma\geq c \ln n$.
    
    For any $\mathbf a \in \mathbf A$, since $u_i(\mathbf a):= v_i(\mathbf{a}_{G(i)  \cup \{i\}})$ and the values $v_i$ are independent across agents, the event that $\mathbf a$ is a pure $\epsilon$-equilibrium can be written as
    \[
    I_{\mathbf a} = \prod_{i \in [n]} \mathbbm 1 \left[  \left\{ v_i(a_i,\mathbf{a}_{G(i)}) \geq \max_{x \in A_i}v_i(x,\mathbf{a}_{G(i)}) -\epsilon \right\} \right] .
    \]
    It follows that if $\mathbf a_{G(i)}\neq\mathbf a'_{G(i)}$ for each $i\in[n]$ then $I_{\mathbf a}$ and $I_{\mathbf a'}$ depend on disjoint sets of the independent values $v_i(\cdot)$.
    This is equivalent to the statement that \emph{if} $G(\Delta(\mathbf a,\mathbf a')) =[n]$ \emph{then} $I_{\mathbf a}$ and $I_{\mathbf a'}$ depend on disjoint sets of values (because, since $G$ is undirected, $\mathbf a_{G(i)}\neq\mathbf a'_{G(i)}$ is equivalent to $i \in \cup_{j \in \Delta(\mathbf a, \mathbf a')} G(j) = G(\Delta(\mathbf a,\mathbf a'))\,$). Now take any $\mathbf a$ and $\mathbf a'$ such that $| \Delta(\mathbf a,\mathbf a') | \geq \lceil n/\gamma \rceil$. Then we can choose $S \subseteq \Delta(\mathbf a,\mathbf a')$ such that $|S| = \lceil n/\gamma \rceil$. Since $G$ is a $\gamma$-expander, the expansion property implies that $| G(\Delta(\mathbf a,\mathbf a')) | \geq | G(S)| \geq \min \{n, \gamma \lceil n/\gamma \rceil\} = n$. This implies that $G(\Delta(\mathbf a,\mathbf a'))=[n]$, and therefore that $I_{\mathbf a}$ and $I_{\mathbf a'}$ depend on disjoint sets of values. Consequently, the two families in \ref{hyp:range} with $r=\lceil n/\gamma\rceil-1$ are functions of disjoint sets of independent values, and hence are independent. And, since $\gamma \geq c \ln n$, we can satisfy \ref{hyp:range} by setting $r_{n,\m}:= \lceil n/c \ln n \rceil - 1 \geq \lceil n/\gamma \rceil - 1$ (which is at least 1 for all sufficiently large $n$). By \cref{lem:ball-power} we therefore obtain
    \[
    b_{n,\m}:=\beta_{\mathbf A}(r_{n,\m}) \leq \left( \frac{e n M}{r_{n,\m}}\right)^{r_{n,\m}}  .
    \]

    We satisfy \ref{hyp:homog} because $\mathbb E[I_{\mathbf a}] = \prod_{i \in [n]} \xi_{F,\alpha_i}^\epsilon := p_{n,\m}$. And, moreover, $\mu_{n,\m}= |\mathbf A|\, p_{n,\m} = \prod_{i \in [n]} \alpha_i \xi_{F,\alpha_i}^\epsilon$.

    We now turn to \ref{hyp:pair}. Consider any $\mathbf a$ and $\mathbf a' \in B_{r_{n,\m}}(\mathbf a) \setminus \{\mathbf a\}$. Partition $[n]$ into $X := G(\Delta(\mathbf a, \mathbf a'))$ and $Y:= [n] \setminus X$ so that $I_{\mathbf a}^X = \prod_{i \in X} \mathbbm 1 \{\dots\}$, $I_{\mathbf a}^Y = \prod_{i \in Y} \mathbbm 1 \{\dots\}$, and $I_{\mathbf a} = I_{\mathbf a}^X I_{\mathbf a}^Y$, and similarly for $\mathbf a'$. By independence across $X$ and $Y$ we can write
    \[
    \mathbb E\bigl[I_{\mathbf a}I_{\mathbf a'}\bigr] =\mathbb E\bigl[I^X_{\mathbf a}I^X_{\mathbf a'}\bigr]\cdot \mathbb E\bigl[I^Y_{\mathbf a}I^Y_{\mathbf a'}\bigr].
    \]
    Since $\mathbf a_{G(i)}\ne\mathbf a'_{G(i)}$ for each $i \in X$, $I^X_{\mathbf a}$ and $I^X_{\mathbf a'}$ are independent. Therefore $\mathbb E\bigl[I^X_{\mathbf a}I^X_{\mathbf a'}\bigr] = \prod_{i \in X} ( \xi_{F,\alpha_i}^\epsilon)^2$. For $Y$, since $I^Y_{\mathbf a}I^Y_{\mathbf a'} \leq I^Y_{\mathbf a}$, we have $\mathbb E\bigl[I^Y_{\mathbf a}I^Y_{\mathbf a'}\bigr] \leq \mathbb E\bigl[I^Y_{\mathbf a}\bigr]=\prod_{i \in Y}\xi_{F,\alpha_i}^\epsilon$. Hence,
    \[
    \mathbb E\bigl[I_{\mathbf a}I_{\mathbf a'}\bigr] \leq \prod_{i \in X} ( \xi_{F,\alpha_i}^\epsilon)^2 \cdot \prod_{i \in Y}\xi_{F,\alpha_i}^\epsilon = p_{n,\m} \prod_{i \in X}\xi_{F,\alpha_i}^\epsilon \leq  p_{n,\m} \cdot (\xi_{F,2}^\epsilon)^{|X|} ,
    \]
    where the last step follows from \cref{lem:xi-cont}. Note that $|X| \geq \min \{n ,\gamma d_H(\mathbf a, \mathbf a')\}$: either $d_H(\mathbf a, \mathbf a')=|\Delta(\mathbf a, \mathbf a')| \geq \lceil n/\gamma \rceil$ in which case, by an argument given above, $|X|= |G(\Delta(\mathbf a, \mathbf a'))| \geq n \geq \min \{n ,\gamma d_H(\mathbf a, \mathbf a')\}$; or $d_H(\mathbf a, \mathbf a')=|\Delta(\mathbf a, \mathbf a')| < \lceil n/\gamma \rceil$ in which case the condition is simply the expander property applied to $\Delta(\mathbf a, \mathbf a')$. Hence, if $d_H(\mathbf a, \mathbf a') \leq r_{n,\m} < n/(c \ln n)$ then $d_H(\mathbf a, \mathbf a') \,c \ln n < n$ and since $\gamma \geq c \ln n$, 
    \[
    |X| \geq \min\{n,\gamma  d_H(\mathbf a, \mathbf a')\} \geq \min\{n,\ d_H(\mathbf a, \mathbf a')\,c\ln n\} = d_H(\mathbf a, \mathbf a') \,c\ln n .
    \]
    We therefore conclude
    \begin{align*}
    \sum_{\mathbf a'\in B_{r_{n,\m}}(\mathbf a)\setminus\{\mathbf a\}} \mathbb E\bigl[I_{\mathbf a}I_{\mathbf a'}\bigr] &= \sum_{k=1}^{r_{n,\m}}\sum_{\mathbf a': d_H(\mathbf a, \mathbf a')=k} \mathbb E\bigl[I_{\mathbf a}I_{\mathbf a'}\bigr] \leq \sum_{k=1}^{r_{n,\m}}\sum_{\mathbf a': d_H(\mathbf a, \mathbf a')=k} p_{n,\m} \, (\xi_{F,2}^\epsilon)^{k\,c\ln n} \\
    &= p_{n,\m} \, \sum_{k=1}^{r_{n,\m}}\sum_{\mathbf a': d_H(\mathbf a, \mathbf a')=k}   n^{- k \,c \ln(1/\xi_{F,2}^\epsilon)} \leq p_{n,\m} \, \sum_{k=1}^{\infty} \frac{(nM)^k}{k!}  n^{- k \,c \ln(1/\xi_{F,2}^\epsilon)} \\&= p_{n,\m} (e^{x_n}-1)
    \end{align*}
    where the first inequality uses the lower bound on $|X|$, the second inequality uses \cref{lem:ball-power} and lets the sum go to infinity, and the final equality uses the definition of $e$, namely, $e^c = \sum_{k\geq 0} \frac{c^k}{k!}$ with $x_n:=M \,n^{\,1-c\ln(1/\xi_{F,2}^\epsilon)}$. Hence \ref{hyp:pair} holds with $\theta_{n,\m}= e^{x_n}-1$, and this completes the proof of Part (a).

    For Part (b), we first record a bound on $b_{n,\m}$. Since $t\mapsto(enM/t)^{t}$ is increasing for $t<nM$ and $r_{n,\m}<n/(c\ln n)$,
    \[
    b_{n,\m}\ \le\ \Bigl(\frac{enM}{r_{n,\m}}\Bigr)^{r_{n,\m}} \leq \bigl(eMc\ln n\bigr)^{n/(c\ln n)}=e^{n\delta_n} \text{ where }
    \delta_n:=\frac{\ln\bigl(eMc\ln n\bigr)}{c\ln n} ,
    \]
    and note that $\delta_n\Longrightarrow0$.

    Suppose first that $\epsilon=0$ and $c>1/\ln 2$. By \cref{lem:xi-cont}, for any $i$, $\xi_{F,\alpha_i}^0 = 1/\alpha_i$, hence $\mu_{n,\m}=1$ and $x_n:=M \,n^{\,1-c\ln(1/\xi_{F,2}^0)} =M \,n^{\,1-c\ln 2} \Longrightarrow 0$. Therefore $\theta_{n,\m} = e^{x_n} - 1\Longrightarrow 0$. Hence,
    \[
    \mu_{n,\m}\bigl(b_{n,\m}p_{n,\m}+\theta_{n,\m}\bigr)=p_{n,\m} b_{n,\m} + \theta_{n,\m} \leq e^{n\delta_n}2^{-n} + \theta_{n,\m}\Longrightarrow0,
    \]
    so the hypotheses of the \cref{thm:master} Part \ref{thm:master:poisson} are satisfied.

    Suppose instead that $\epsilon>0$ and $c>1/\ln\bigl(1/\xi^{\epsilon}_{F,2}\bigr)$. Since $\epsilon >0$, by \cref{lem:xi-cont}, $\tau:=\min_{2\le\alpha\le M}\alpha\,\xi^{\epsilon}_{F,\alpha}>1$, and therefore $\mu_{n,\m} \geq \tau^n \Longrightarrow \infty$. And, since $c>1/\ln\bigl(1/\xi^{\epsilon}_{F,2}\bigr)$, we have $x_n=M\,n^{\,1-c\ln(1/\xi^{\epsilon}_{F,2})}\Longrightarrow0$ and hence $\theta_{n,\m}\Longrightarrow0$; combined with $\mu_{n,\m}\Longrightarrow\infty$ this gives $\theta_{n,\m}/\mu_{n,\m}\Longrightarrow0$, so the hypotheses of the \cref{thm:master} Part \ref{thm:master:existence} are satisfied. For the \cref{thm:master} Part \ref{thm:master:clt}, observe that $c>1/\ln\bigl(1/\xi^{\epsilon}_{F,2}\bigr)$ implies that $\xi^{\epsilon}_{F,2}<1$ and therefore $b_{n,\m}\,p_{n,\m} \leq e^{\,n\delta_n} \, (\xi^{\epsilon}_{F,2})^n \Longrightarrow\ 0 $. Finally,
    \[
    \frac{b_{n,\m}^{2}}{\sqrt{\mu_{n,\m}}} \leq \frac{e^{2n\delta_n}}{\tau^{\,n/2}}=e^{n\bigl(2\delta_n-\frac12\ln\tau\bigr)}\ \Longrightarrow\ 0 .
    \]
    So the hypotheses of the \cref{thm:master} Part
    \ref{thm:master:clt} are satisfied (by \cref{cor:master-clt-suff}), which completes the proof.
\end{proof}

\subsection{\texorpdfstring{Proof of \cref{thm:independent_discrete}}{Proof for the independent, discrete case}}\label{sec:thm-ind-disc-grid}

\begin{proposition}[Independent, discrete case]\label{prop:grid-disc}
Fix an integer $M\ge2$. For each $n\ge2$ and each $\m = (\mathbf A,\rho)\in\mathcal I(n) = \mathcal A_M(n) \times \mathcal R(n)$ let $I_{\mathbf a}$ be the indicator that $\mathbf a$ is a pure $\epsilon$-equilibrium of $g=(\mathbf A, \{u_i\}_{i \in [n]})$, where $g$ has utilities drawn from $\mathbb J_{n,\m}$ as described in \cref{sec:independent_discrete}. Then we have:
\begin{enumerate}[(a),topsep=0ex,itemsep=0ex,leftmargin=0cm]
\item $(I_{\mathbf a})_{\mathbf a\in\mathbf A}$ is a $(r_{n,\m},\theta_{n,\m})$-well-behaved indicator grid with
\[
r_{n,\m}=1 \text{ and }  \theta_{n,\m}:=nM^2\,\prod_{i\in[n]}\xi_{\rho,\alpha_i} .
\]
Moreover, letting $S_{n,\m}:=\sum_{\mathbf a\in\mathbf A}I_{\mathbf a}$, $\mu_{n,\m}:=\mathbb E[S_{n,\m}]$, and $p_{n,\m}:=\frac{\mu_{n,\m}}{|\mathbf A|}$, we have
\[
\mu_{n,\m}=\prod_{i\in[n]}\alpha_i \,\xi_{\rho,\alpha_i} ,  \text{ and } b_{n,\m}:= \beta_{\mathbf A}(r_{n,\m}) \leq nM.
\]
\item If $n/\rho \Longrightarrow 0$ then the hypotheses of the \cref{thm:master} Part \ref{thm:master:poisson} are satisfied; and if $n/\rho \Longrightarrow \infty$ then the hypotheses of the \cref{thm:master} Parts \ref{thm:master:clt} and \ref{thm:master:existence} are satisfied.
\end{enumerate}
\end{proposition}
\noindent\cref{thm:independent_discrete} follows directly from \cref{prop:grid-disc} and the \cref{thm:master}.\hfill$\Box$

Observe that the indicator grids induced by the independence-continuous model of \cref{sec:independent_continuous} and by the independent-discrete model of \cref{sec:independent_discrete} are identical up to $\xi$. This hints at a formal equivalence between the models. In fact, if we let $\epsilon$ vary with $n$ and $\m$ in the continuous model, then we can replicate the outcomes of the discrete model, as stated in the remark below.

\begin{remark*}
Fix an integer $n\ge2$, a grid $\mathbf A\in\mathcal A_M(n)$, and a resolution $\rho\in\mathcal R(n)$, and assume that $F$ is continuous with support an interval of length $\ell<\infty$. Then there is a unique $\epsilon^\ast=\epsilon^\ast(n,\mathbf A,\rho)\in[0,\ell]$ with\footnote{This follows from the intermediate value theorem: the left-hand side is continuous in $\epsilon^\ast$ and strictly increasing from $1$ to $\prod_{i\in[n]}\alpha_i$ as $\epsilon^\ast$ goes from $0$ to $\ell$, and the right-hand side lies in $[1,\prod_{i\in[n]}\alpha_i]$.}
\[
\prod_{i\in[n]}\alpha_i\,\xi^{\epsilon^\ast}_{F,\alpha_i}
\;=\;
\prod_{i\in[n]}\alpha_i\,\xi_{\rho,\alpha_i} .
\]
That is, at $\epsilon=\epsilon^\ast$ the two models induce indicator grids with the same dependence range $r$, ball size $b$, per-profile probability $p$, local pair-correlation parameter $\theta$, and mean $\mu$; these are the only quantities entering the conditions verified in \cref{prop:grid-cont} and \cref{prop:grid-disc}, so the two models satisfy the hypotheses of the \cref{thm:master} under identical circumstances and the same asymptotic conclusions follow. As an example, one may verify that if $\alpha_i=\alpha$ for every $i$ and $F$ is the cdf of a uniform random variable on an interval of length $1$, then $\epsilon^\ast=\tfrac{1}{2\rho}+h$ where $h$ is a term of order at most $\rho^{-2}$.
\end{remark*}

\begin{proof}[Proof of \cref{prop:grid-disc}]
    The proof of the first part follows verbatim from the argument we gave to prove Part (a) of \cref{prop:grid-cont} with $\epsilon=0$, with $\xi_{F,\alpha}^\epsilon$ replaced by $\xi_{\rho,\alpha}$, and with \cref{lem:xi-cont} replaced by \cref{lem:xi-disc}.

    We therefore turn directly to the proof of Part (b). By \cref{lem:xi-disc}, we have
    \[ 1+\frac{n}{\rho} \leq \prod_{i\in[n]}\Bigl(1+\frac{\alpha_i}{2\rho}\Bigr) \leq \mu_{n,\m} \leq \Bigl(1+\frac{1}{2\rho}\Bigr)^{\sum_{i\in[n]}\alpha_i} \leq \exp\Bigl(\frac{Mn}{2\rho}\Bigr),\]
    where the first inequality is an application of $\alpha_i \geq 2$ and the Weierstrass product inequality and the last inequality uses $\alpha_i \leq M$ and $(1+x)^k \leq e^{xk}$ for $x \geq 0$. It follows that if $n/\rho \Longrightarrow 0$ then $\mu_{n,\m}\Longrightarrow 1$ and $\mu_{n,\m}\bigl(b_{n,\m}p_{n,\m}+\theta_{n,\m}\bigr)  \leq \frac{\mu_{n,\m}^2}{2^n}(nM + nM^2) \Longrightarrow0$, so the hypotheses of the \cref{thm:master} Part \ref{thm:master:poisson} are satisfied. And, if $n/\rho \Longrightarrow \infty$ then $\mu_{n,\m}\Longrightarrow \infty$ and $\theta_{n,\m}/\mu_{n,\m} = nM^2 p_{n,\m} / \mu_{n,\m} \leq \frac{nM^2}{2^n}\Longrightarrow 0$ so the hypotheses of the \cref{thm:master} Part \ref{thm:master:existence} are satisfied; and, moreover, by \cref{lem:xi-disc}, $b_{n,\m}p_{n,\m} \leq nM (3/4)^n \Longrightarrow0$ and $\theta_{n,\m}\leq nM^2(3/4)^n \Longrightarrow 0$, and finally
    \begin{align*}
    &\min\left\{\ \frac{b_{n,\m}^{2}}{\sqrt{\mu_{n,\m}}}\ ,\ \
    \mu_{n,\m}\bigl(b_{n,\m}p_{n,\m}+\theta_{n,\m}\bigr)+\theta_{n,\m}\ \right\} \\
    &\qquad\leq \min\left\{\ \frac{b_{n,\m}^{2}}{\sqrt{\mu_{n,\m}}}\ ,\ \
    \mu_{n,\m}\bigl(b_{n,\m}p_{n,\m}+\theta_{n,\m}\bigr)\ \right\} + \theta_{n,\m} \\&\qquad
    \leq  \left(\frac{b_{n,\m}^{2}}{\sqrt{\mu_{n,\m}}} \right)^{\frac{2}{3}} \Bigl( \mu_{n,\m}\bigl(b_{n,\m}p_{n,\m}+\theta_{n,\m}\bigr) \Bigr)^{\frac{1}{3}} + \theta_{n,\m}\\
    &\qquad= b_{n,\m}^\frac{4}{3} \bigl(b_{n,\m}p_{n,\m}+\theta_{n,\m}\bigr)^\frac{1}{3} + \theta_{n,\m},
    \end{align*}
    where the key step follows from $\min\{x,y\}\leq x^c y^{1-c}$ for $x,y \geq 0$ and $c \in [0,1]$. But $ b_{n,\m}^\frac{4}{3}\bigl(b_{n,\m}p_{n,\m}+\theta_{n,\m}\bigr)^\frac{1}{3} \Longrightarrow 0$ so the hypotheses of the \cref{thm:master} Part \ref{thm:master:clt} are satisfied (by \cref{cor:master-clt-suff}), which completes the proof.
\end{proof}

\singlespacing{
\small
\setlength{\bibsep}{3pt plus 1pt}

}

\end{document}